\documentclass[a4paper,12pt,preprint,amsmath,showpacs,
nofootinbib,superscriptaddress]{revtex4}
\usepackage{mathrsfs}
\usepackage{amssymb}
\usepackage{amsmath}
\usepackage{graphicx}
\usepackage{multirow}
\usepackage{subfigure}
\usepackage{float}
\usepackage{hyperref}
\usepackage{graphics}
\usepackage{color}
\usepackage{bookmark}

\def\be{\begin{equation}}
\def\ee{\end{equation}}
\newcommand{\bea}{\begin{equation} \begin{array}{c}}
\newcommand{\eea}{ \end{array} \end{equation}}

\def\as{\alpha_s}

\def\euv{\epsilon_{UV}}
\def\eir{\epsilon_{IR}}
\def\e{\epsilon}

\newcommand{\beqn}{\begin{eqnarray}}
\newcommand{\eeqn}{\end{eqnarray}}

\newcommand{\nn}{\nonumber}

\newcommand{\gsim}{\lower.7ex\hbox{$\;\stackrel{\textstyle>}{\sim}\;$}}
\newcommand{\lsim}{\lower.7ex\hbox{$\;\stackrel{\textstyle<}{\sim}\;$}}

\def\e{\epsilon}

\begin{document}
\title{Constraints on Randall-Sundrum model from the events of dijet production with QCD next-to-leading order accuracy at the LHC}

\author{Shi Ang Li}
\affiliation{Department of Physics and State Key Laboratory of
Nuclear Physics and Technology, Peking University, Beijing 100871,
China}
\author{Chong Sheng Li}
\email{csli@pku.edu.cn}
\affiliation{Department of Physics and State Key Laboratory of
Nuclear Physics and Technology, Peking University, Beijing 100871,
China}
\affiliation{Center for High Energy Physics, Peking University, Beijing, 100871, China}
\author{Hai Tao Li}
\affiliation{Department of Physics and State Key Laboratory of
Nuclear Physics and Technology, Peking University, Beijing 100871,
China}
\author{Jun Gao}
\affiliation{Department of Physics, Southern Methodist University, Dallas, TX 75275-0181, USA}

\begin{abstract}
We study the dijet production in Randall-Sundrum model at the LHC with QCD next-to-leading(NLO) order accuracy. Our results show that the QCD NLO corrections can increase the total cross sections by more than $80\%$ and reduce the scale dependence. We also explore in detail several important kinematic distributions at the NLO level. Moreover, we discuss the upper limits of the KK graviton excluded mass range and the allowed parameter space for the coupling constant and KK graviton mass, using the experiment data.
\end{abstract}

\bibliographystyle{unsrt}
\maketitle
\flushbottom
\section{INTRODUCTION}\label{s1}

Searching  for new physics is one of the most important tasks at the LHC.  In many extensions of the standard model(SM), there  exist massive particles that couple to quarks or gluons, which may be observed as a narrow resonance in dijet production, such as $W'$, $Z'$, excited quarks, axigluon,  Klauza-Klein(KK) graviton from extra dimensions.  Therefore, the study of dijet events provides a possibility to probe new physics effects. In the SM, the dijet events are mostly produced through quantum chromodynamics(QCD) interactions in hadron colliders, which predicts a smooth and steeply falling dijet mass spectrum. Experiments at the LHC have already used the dijet invariant mass to constrain the mass of these new resonances~\cite{CMS:2012eba,CMS:2012nba,Chatrchyan:2013qha}. Randall-Sundrum(RS) model~\cite{Randall:1999ee,Randall:1999vf} is one among various new physics models which can solve the large hierarchy problem of the weak and the Plank scale.

 In RS model, the extra dimension is assumed to be located on a $S_1/Z_2$ orbifold, which has two fixed points, $\phi\,=\,0$ and $\phi\,=\,\pi$. They correspond to high energy brane and the brane we live on, respectively. Graviton is the only particle that can propagate through the bulk between these two branes. The 5-dimensional warped matric is given by:
\beqn\label{eq:fir}
ds^2\,=\,e^{-2 k r |\phi|}(\eta_{\mu\nu}\,&+&\,\frac{2}{M_P^{3/2}}h_{\mu\nu})dx^{\mu}dx^{\nu}\,-\,r^2 d \phi^2\,,\\\nn
0\,&\leq&\,|\phi|\,\leq\,\pi\, ,
\eeqn
where $\phi$ is the five-dimensional coordinate, $k$ is a scale of order of the Plank scale, $r$ is the compactification radius of the extra dimensional circle, and $h_{\mu\nu}$ is the graviton metric. Solving the 5-dimensional Einstein equation and using Eq.~(\ref{eq:fir}), we can get the relation between the 4-dimensional reduced Plank scale $\bar{M}_{p}$ and the 5-dimensional Plank scale $M_P$~\cite{Randall:1999ee},
\be
\bar{M}_{p}\,=\,\frac{M_P^3}{k}(1\,-\,e^{-2 k r \pi})\,.
\ee
The physical mass $m$ of a field in 4-dimension, is related to the fundamental mass parameter $m_0$ as following:
\be
m\,=\,e^{-k r \pi} m_0\,.
\ee
thus the hierarchy problem can be solved by assuming $k r\,\sim\,12$.

 There also exist KK towers of the massive spin-2 graviton that can interact with the SM fields, and their 4-dimensional effective Lagrangian is given by~\cite{Hewett:1998sn,Davoudiasl:1999jd}:
\be
\mathfrak{L}\,=\,-\frac{1}{\bar{M}_{p}}T^{\alpha\beta}(x)h_{\alpha\beta}^{(0)}(x)\,-\,\frac{1}{\Lambda_{\pi}}T^{\alpha\beta}(x)\sum_{n=1}^{\infty}
h_{\alpha\beta}^{(n)}(x)\, ,
\ee
with
\be\label{l m x1}
\kappa=\frac{1}{\Lambda_{\pi}}\,=\,\frac{1}{\bar{M}_{p}}e^{2 k r \pi}\,=\,\frac{x_1 k}{m_{KK} \bar{M}_{p}}\, ,
\ee
where $\kappa$ stands for the coupling constant between KK graviton and SM particles and $\Lambda_{\pi}$ is around the electroweak scale. $m_{KK}$ is the mass of the 1st KK excitation mode of the graviton, which we will focus on in this paper. $x_1$ is the 1st root of the first order Bessel function. Then the masses of the $1$th KK excitation modes are given by
\be\label{l m xn}
m_{KK}\,=\,k x_1 e^{-k r \pi}\,=\,\frac{k}{ \bar{M}_{p}}\frac{x_1}{\kappa}\,,
\ee
From Eq.~(\ref{l m x1}) and Eq.~(\ref{l m xn}), the graviton sector of the RS model is completely determined by two parameters $m_{KK}$ and $k/\bar{M}_{p}$.

The RS KK graviton can be produced through both the $gg$ fusion and the $q\bar{q}$ annihilation at the LO. The detailed Feynman rules of the graviton couplings can be found in Ref.~\cite{Han:1998sg}, and the propagator for the massive spin-2 KK states is~\cite{Mathews:2005bw}
\begin{equation}
P_{\mu\nu,\rho\sigma}^{G}(k)
\ =\ {\frac{i}{2}\frac{ B_{\mu\nu,\rho\sigma}(k)}{
 k^2-m^2_{KK}+i m_{KK} \Gamma_{KK}}}\ ,
\label{prop}
\end{equation}
where
\begin{eqnarray}
B_{\mu\nu,\rho\sigma}(k) &=&
\left(\eta_{\mu\rho}-{\frac{k_\mu k_\rho}{ m_{KK}^2}}\right)
\left(\eta_{\nu\sigma}-{\frac{k_\nu k_\sigma}{ m_{KK}^2}}\right)
+\left(\eta_{\mu\sigma}-{\frac{k_\mu k_\sigma}{ m_{KK}^2}}\right)
\left(\eta_{\nu\rho}-{\frac{k_\nu k_\rho}{ m_{KK}^2}}\right)\nonumber\\
&& - {\frac{2}{{n-1}}}\left(\eta_{\mu\nu}-{\frac{k_\mu k_\nu}{ m_{KK}^2}}\right)
\left(\eta_{\rho\sigma}-{\frac{k_\rho k_\sigma}{ m_{KK}^2}}\right)\ .
\label{B}
\end{eqnarray}
where $\Gamma_{KK}$ is the width of the heavy resonance, respectively.

The LO cross section and the signal for dijet production via KK graviton exchange have been calculated in the RS model in Refs.~\cite{Atwood:1999qd,Allanach:2002gn}. To put more stringent bound on the parameters of the model at the LHC, we need the QCD NLO corrections to promote the theoretical accuracy. Presently, many processes are available for NLO accuracy, including single KK graviton production\cite{Mathews:2005bw,Li:2006yv} and graviton decay to different final states such as Drell-Yan~\cite{Mathews:2004xp,Kumar:2006id}, di-photon~\cite{Kumar:2009nn,Kumar:2008pk}, $Z\,Z$~\cite{Agarwal:2009xr,Agarwal:2009zg}, $W^+ \,W^-$~\cite{Agarwal:2010sp,Agarwal:2010sn}, $Z$+missing energy~\cite{Chen:2014oha} and $t\,\bar{t}$~\cite{Gao:2010bb}. Since K factors at the NLO level in these processes are large, it is also essential to go beyond LO for dijet final state process. In this paper, we present a QCD NLO calculation to the KK-graviton production and decay in the dijet channel at the LHC, and give constraints on the relative parameters with NLO accuracy through comparing with the latest dijet event data from the CMS collaboration~\cite{Chatrchyan:2013qha}.

This paper is organized as follows. In Sec.~\ref{s3} we show the analytic results for the LO and QCD NLO cross sections and the consistent treatment for including the QCD NLO effects of KK graviton decay width. In Sec.~\ref{s5} we present the numerical predictions for inclusive and differential cross sections at the LHC. We simulate the signal for RS KK graviton at the LHC and update the constraints on the KK graviton mass using recent measurement with the NLO results. Some of the lengthy analytic expressions are summarized in Appendix.

\section{ANALYTICAL Results}\label{s3}

In this section, we present the analytical results for dijet production via KK graviton exchange. The QCD NLO corrections can be factorized into two independent gauge invariant parts, i.e.,  KK graviton produced at the NLO with a subsequent decay at the LO, and produced at the LO with a subsequent decay at the NLO, similar to the cases of Refs.~\cite{Cao:2004ap,Gao:2010bb}. We neglect interference between radiation in the two stages, which are expected to be small, of order $\mathcal{O}(\alpha_s\Gamma_{KK}/M_{KK})$ ~\cite{Fadin:1993kt,Fadin:1993dz,Melnikov:1993np}. This whole procedure can be illustrated as follows:
\beqn
|\mathcal{M}^{tree}_{2\rightarrow 2}|^2&=&|\mathcal{M}^{tree}_{pro}|^2\otimes|\mathcal{M}^{tree}_{dec}|^2\otimes|P_{G}|^2\,,\\\nn
|\mathcal{M}^{real}_{2\rightarrow3}|^2&=&\{|\mathcal{M}^{tree}_{pro}|^2\otimes|\mathcal{M}^{real}_{dec}|^2+|\mathcal{M}^{real}_{pro}|^2\otimes|\mathcal{M}^{tree}_{dec}|^2\}\otimes|P_{G}|^2\,,\\\nn
\mathcal{M}^{tree\ast}_{2\rightarrow 2}\mathcal{M}^{loop}_{2\rightarrow 2}&=&\{|\mathcal{M}^{tree}_{pro}|^2\otimes(\mathcal{M}^{tree\ast}_{dec}\mathcal{M}^{loop}_{dec})+|\mathcal{M}^{tree}_{dec}|^2\otimes(\mathcal{M}^{tree\ast}_{pro}\mathcal{M}^{loop}_{pro})\}\otimes|P_{G}|^2\,,\nn
\eeqn
where we have suppressed the possible Lorentz indices here for simplicity.

\subsection{Leading Order Results}
\begin{figure}[ht]
\begin{center}
  \includegraphics[width=1.0\textwidth]{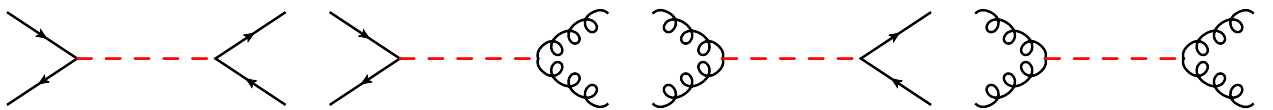}\\
  \caption{ \label{fig:tree} Tree-level Feynman diagrams for KK graviton production and decay into dijet.}
\end{center}
\end{figure}
 The LO Feynman diagrams for the production and decay of the KK graviton are shown in Fig.~\ref{fig:tree}.
After summing over spin and color of the final state particles and averaging over spin and color of the initial states,  the amplitude squares are
\begin{eqnarray}
\overline{|\mathcal{M}^{tree}_{q \bar{q}\rightarrow q \bar{q}}|}^2\,&=&\,\frac{1}{512}\,\kappa^4\,(s^4\,+\,10 \,s^3\, t\,+\,42\, s^2\, t^2\,+\,64\, s \,t^3\,+\,32 \,t^4)R(s)\,\\
\overline{|\mathcal{M}^{tree}_{g g \rightarrow q \bar{q}}|}^2\,&=&\,-\frac{3}{256}\,\kappa^4\, t\,(\,s^2\,+\,2\,s\,t\,+\,2\,t^2\,)\,(\,s\,+\,t\,)R(s) \,\\
\overline{|\mathcal{M}^{tree}_{ q \bar{q} \rightarrow g g}|}^2\,&=&\,-\frac{1}{24}\,\kappa^4\, t\,(\,s^2\,+\,2\,s\,t\,+\,2\,t^2\,)\,(\,s\,+\,t\,)R(s) \,\\
\overline{|\mathcal{M}^{tree}_{g g \rightarrow g g}|}^2\,&=&\,\frac{1}{64}\,\kappa^4\,( \,s^4\,+\,4\,s^3 \,t\,+\,6\, s^2\, t^2\,+\,4\, s\, t^3\,+\,2\, t^4\,)R(s) \,
\end{eqnarray}
where the Mandelstam variables $s,t,u$ are defined as,
 \begin{equation}
 s=(p_1+p_2)^2,\,t=(p_1-p_3)^2,\,u=(p_1-p_4)^2\, .
 \end{equation}
$R(s)$ represents the LO contribution from propagator for Breit-Wigner resonance, which can be written as
\be
R(s)=\frac{1}{(s-m_{KK}^2)^2+\Gamma_{KK}^2m_{KK}^2}\, ,
\label{eq:br}
\ee
Throughout this paper, we work in  the 't Hooft-Feynman gauge.

 At hadron colliders, the LO total cross section is obtained by  convoluting the partonic cross section  with the parton distribution functions, which is
 \begin{align}
\sigma(p\,p\rightarrow j\,j) &= \sum_{q\bar{q}}\sum_{ab}\int dx_1 dx_2 [G_{q/p}(x_1,\mu_f)G_{\bar{q}/p}(x_2,\mu_f)\hat{\sigma}_{q\bar{q}\to ab }+(x_1\leftrightarrow x_2)]
\nn \\ &
  + \sum_{ab}\int dx_1 dx_2 G_{g/p}(x_1,\mu_f)G_{g/p}(x_2,\mu_f)\hat{\sigma}_{gg\to ab}\,,
\end{align}
where $\mu_f$ is the factorization scale.
The LO partonic cross section is defined as
\beqn
\hat{\sigma}^B_{ij\to ab}&=&\frac{1}{2s}\int d{\rm PS}_2 \overline{|\mathcal{M}^{LO}_{i j\rightarrow ab}|}^2 \,.
\label{eq:fact}
\eeqn

\subsection{NEXT-TO-LEADING ORDER QCD CORRECTIONS}
\subsubsection{Virtual Corrections}

\begin{figure}[ht]
\begin{center}
  \includegraphics[width=1.0\textwidth]{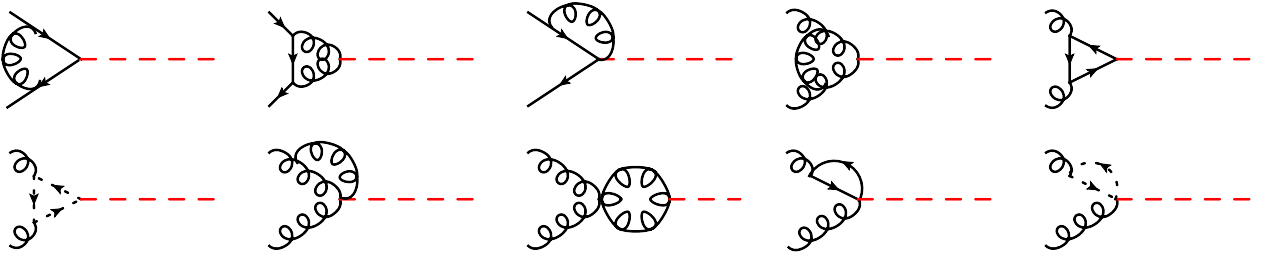}\\
  \caption{ \label{fig:virtual} One-loop Feynman diagrams for the production of KK graviton.   }
\end{center}
\end{figure}

 The loop diagrams for the production part are shown in Fig.~\ref{fig:virtual}. The virtual corrections contain both UV and IR divergences, with the UV divergences renormalized by introducing counterterms. Using the on-shell subtraction scheme, we define all the renormalization constants for massless quarks and gluons, which are given by

\begin{align}
\delta Z_q^{OS}\,&=-\frac{\as}{3\pi}C_\e\left\{\frac{1}{\euv}-\frac{1}{\eir}\right\}\, ,
\nn \\
\delta Z_{G}^{OS}&=-\frac{\alpha_S}{2\pi} \left(\frac{n_f}{3}-\frac{5}{2}\right) C_\e\left\{\,\frac{1}{\epsilon_{UV}}\,-\,\frac{1}{\epsilon_{IR}}\,\right\} -\frac{\alpha_S}{6\pi}C_\e\left(\frac{1}{\e_{UV}}\right)\, ,
\end{align}
where $C_\e=\Gamma(1+\e)(4\pi\mu_r^2/m_t^2)^\e$ and $n_f=5$ is the number of flavors of the massless quarks and $\mu_r$ is the renormalization scale.
 For the $q\bar{q}$ initial states, the renormalized virtual corrections to partonic cross section are
\begin{align}
\label{eq:MVqq}
\hat{\sigma}^{V}_{q \bar{q}}&=\hat{\sigma}^{B}_{q\bar{q}}\,\frac{\alpha_s}{2\pi}\,D_\epsilon\,
\left\{\frac{A_2^{v,q}}{\eir^2}+\frac{A_1^{v,q}}{\eir}+A_0^{v,q}\right\}\, ,
\end{align}
with
\beqn
D_\epsilon&=&\frac{\Gamma(1-\e)}{\Gamma(1-2\e)}\left(\frac{4\pi\mu_r^2}{s}\right)^\e\\\nn
A_2^{v,q}&=&-\frac{8}{3},\\\nn
A_1^{v,q}&=&-4,\\\nn
A_0^{v,q}&=&\frac{8}{9}(\pi^2-15).
\eeqn
For the gluon initial states,  the renormalized virtual corrections are
\be
\label{eq:MVgg}
\hat{\sigma}^{V}_{g g}=\hat{\sigma}^{B}_{g g}\,\frac{\alpha_s}{2\pi}\,D_\epsilon\,
\left\{\frac{A_2^{v,g}}{\eir^2}+\frac{A_1^{v,g}}{\eir}+A_0^{v,g}\right\}\, .
\ee
with
\beqn
A_2^{v,g}\,&=&\, -6\, , \nn \\
A_1^{v,g}\,&=&\,\frac{2 n_f - 33}{3}\, ,\\\nn
A_0^{v,g}\,&=&\,\frac{1}{18}(35 n_f+36 \pi^2-609)+\frac{1}{18 s}\{12 m_t^2(6 C_0 m_t^2+3C_0 s+11)\\\nn
           &-&12(5 m_t^2 + s)\left[\ln\left(\frac{\mu_r^2}{m_t^2}\right)+\ln\left(\frac{\mu_r^2}{s^2}\right)\right]+47s\}.
\eeqn
where $C_0$ is the finite scalar integral in Ref.~\cite{Beenakker:1988jr}, which shows as
\beqn
C_0 (0,0,s;m^2,m^2,m^2)&=&\frac{x_s}{m^2 (1-x_s^2)}[-\frac{1}{2}\ln^2 x_s + 2\ln(x_s)\ln(1+x_s)\\\nn
                       &+&2{\rm Sp}(-x_s)+\frac{\pi^2}{6}]\, ,
\eeqn
with
\beqn
{\rm Sp}(z)&=&\int_0^1 dt \frac{\ln(1-z t)}{t}\, , \nn \\
x_s\;\;&=&-\textit{K}(s+i\epsilon,m_t,m_t)\, , \nn  \\
K(z,m,m^\prime) &= &\frac{1-\sqrt{1-4 m m^\prime /[z-(m-m^\prime)^2]}}{1+\sqrt{1-4 m m^\prime /[z-(m-m^\prime)^2]}}
\;\;\;\;\;\;\;\;\;\;\;\; z \neq (m-m^\prime)^2 \, ,\nn \\
K(z,m,m^\prime) &= &-1\;\;\;\;\;\;\;\;\;\;\;\;\;\;\;\;\;\;\;\;\;\;\;\;\;\;\;\;\;\;\;\;\;\;\;\;\;\;\;\;\;\;\;\;\;\;\;\;\;\;\;\;\;\;\;\;\;\;z = (m-m^\prime)^2\,.
\eeqn

Note that in above renormalized amplitudes, all the UV divergences cancel each other, leaving the remaining IR divergences and the finite terms.

\subsubsection{Real corrections}
The real corrections consist of radiation of an additional gluon, or massless quark (anti-quark) in the final state. For real particle emission, the phase space integration contains both soft and collinear singularities. We adopt the two-cutoff phase space slicing method~\cite{Harris:2001sx} to isolate all the IR singularities, where the phase space is divided into different regions by introducing two small cutoffs $\delta_s$ and $\delta_c$. The soft cutoff $\delta_s$ separates the phase space into the soft region and hard region according to the soft condition $E_i\leq\delta_s s$, which can be written as
\be
\hat{\sigma}^R_{ij}\,=\,\hat{\sigma}^S_{ij}\,+\,\hat{\sigma}^H_{ij}\, ,
\ee
Furthermore, the hard piece is divided into two regions by collinear cutoff $\delta_c$ according to the collinear condition $-\delta_c s<(p_i-p_{5})^2<0$,
\be
\hat{\sigma}^H_{ij}\,=\,\hat{\sigma}^{HC}_{ij}\,+\,\hat{\sigma}^{\overline{HC}}_{ij}\, .
\ee

 The $\hat{\sigma}^{HC}_{ij}$ contains the collinear divergences, which can be obtained by integration over the phase space of the emitted partons. The hard non-collinear part $\hat{\sigma}^{\overline{HC}}_{ij}$ is finite, and we can compute it using standard Monte Carlo integration techniques.

\paragraph{\normalsize{\textbf{Soft gluon emission}}}

In the limit that the energy of the emitted gluon becomes small, i.e. $E_5\leq \delta_s\sqrt{s} /2$, the amplitude square can be factorized into the Born amplitudes times an eikonal factor $\Phi_{eik}$
\beqn
\overline{\sum}|M_{\rm real}(1+2\rightarrow 3+4+5)|^2_{\rm soft}\rightarrow(4\pi\alpha_s)\sum\overline{|M_0|}^2\Phi_{\rm eik}\, ,
\eeqn
with
\be
\Phi_{\rm eik}=C_I\,\frac{s}{p_1\cdot p_5 p_{2}\cdot p_5}\, ,
\label{eq:eiko}
\ee
where $C_I=C_F$ for $q\bar{q}$ initial state and $C_I=C_A$ for $gg$ initial state. Here we only consider the situation for the initial state. Then the parton level cross section in the soft region can be expressed as
\be
\hat{\sigma}^{S}_{ij}=\frac{1}{2s}\int \overline{|M_{\rm real}|}^2|_{\rm soft} d\Gamma_3^{\rm soft}\, ,
\label{eq:socr}
\ee
where $d\Gamma_3^{\rm soft}$ is the three-body phase space in the soft region, which can be factorized :
\be
\label{eq:soph}
d\Gamma|_3^{\rm soft}=d\Gamma_2\left[\left(\frac{4\pi}{s}\right)^{\epsilon}\frac{\Gamma(1-\e)}{\Gamma(1-2\e)}\frac{1}{2(2\pi)^2}\right]dS\, ,
\ee
with
\be
dS=\frac{1}{\pi}\left(\frac{4}{s}\right)^{-\e}\int_0^{\delta_s \sqrt{s}/2}dE_5 E_5^{1-2\e}sin^{1-2\e}\theta_1 d\theta_1 sin^{-2\e}\theta_2 d\theta_2\,.
\ee

After the integration over the soft gluon phase space, we have
\be
\hat{\sigma}_{ij}^S=\frac{\alpha_s}{2\pi}\hat{\sigma}^B_{ij}D_{\e}\left(\frac{A_2^s}{\e ^2}+\frac{A_1^s}{\e}+A_0^s\right)\,,
\ee
with
\beqn
A_2^s&=&2C_I\,,\\\nn
A_1^s&=&-4C_I \ln \delta_s\,,\\\nn
A_0^s&=&4C_I \ln^2 \delta_s\,.\nn
\eeqn
For soft gluon radiated from outgoing partons, it gives the same results. Here we do not show their expressions.

\paragraph{\normalsize{\textbf{Collinear emission}}}
In this section we discuss the collinear singularities in $\sigma_{HC}$, which is treated differently according to whether the singularities are from initial or final state.
\subparagraph{Initial state collinear radiation}

\begin{figure}[ht]
\begin{center}
  \includegraphics[width=1.0\textwidth]{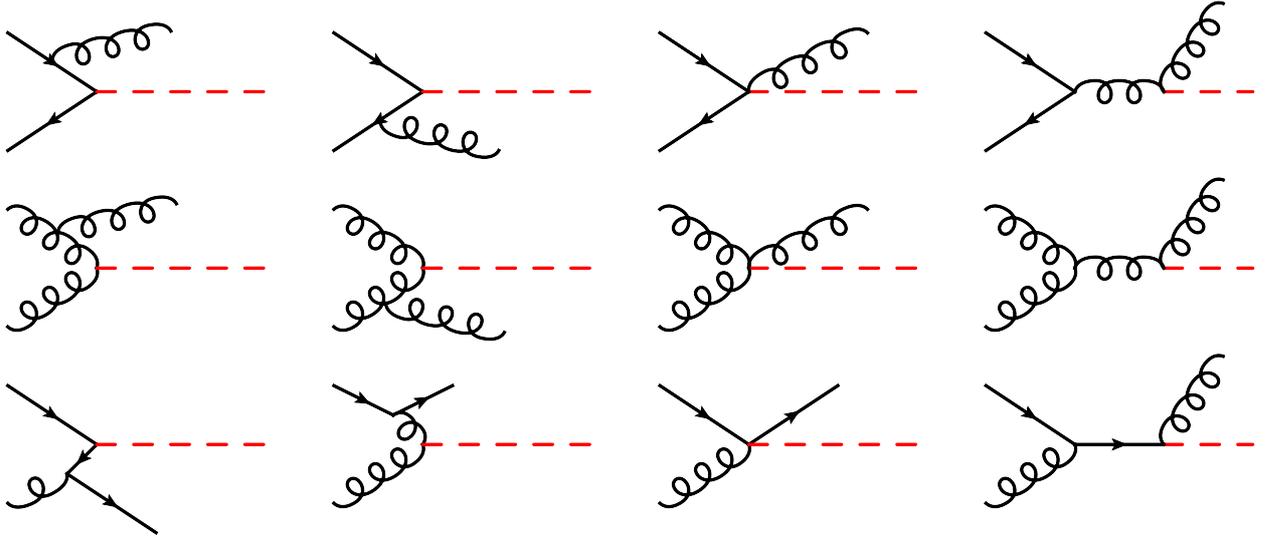}\\
  \caption{ \label{fig:ini_real} Real correction Feynman diagrams for the production of the KK graviton.}
\end{center}
\end{figure}
The real emission diagrams from initial states are shown in Fig.~\ref{fig:ini_real}. In the hard collinear region, $E_5>\delta_s \sqrt{s}/2$ and $0<-t_{i5}<\delta_c s$, the emitted hard gluon(quark) is collinear to one of the incoming partons.
As a consequence of the factorization theorem, the matrix element square can be factorized into the product of the born amplitude square and the Altarelli-Parisi splitting functions $P_{ij}(z,\e)$~\cite{Altarelli:1977zs}
\be
\overline{\sum}|M_3(1+2\rightarrow 3+4+5)|^2_{\rm coll}\rightarrow(4\pi\alpha_s\mu_r^{2\e})\sum\overline{|M_0|}^2\left[\frac{-2P_{1'1}(z,\e)}{z t_{15}}+\frac{-2P_{2'2}(z,\e)}{z t_{25}}\right]\,.
\ee
Here $z$ denotes the fraction of the momentum of $1(2)$ carried by parton $1'(2')$ with the emitted parton $5$ taking a fraction $(1-z)$.

Moreover, the collinear three body final phase space can be factorized in the collinear limit. For example, in the limit $0<-t_{15}<\delta_c s$, it has the following form~\cite{Harris:2001sx}
\be
d\Gamma_3(1+2\rightarrow3+4+5)|_{ \rm coll}\rightarrow d\Gamma(1'+2\rightarrow 3+4)|_{s'=z s}\frac{(4\pi)^{\e}}{16\pi^2 \Gamma(1-\e)}d z d t_{15}[-(1-z)t_{15}]^{-\e}\,.
\ee
Substituting the matrix elements square and phase space in collinear limits into the hard collinear cross section, we have
\beqn\label{eq:hc}
d\sigma_{ij}^{HC}&=&\frac{\alpha_s}{2\pi}D_{\e}(-\frac{1}{\e})\delta_c^{-\e}\{d\hat{\sigma}_{q\bar{q}}^B [P_{1'1}(z,\e)G_{1/p}(x_1/z)G_{2/p}(x_2)\\\nn
&+&P_{2'2}(z,\e)G_{2/p}(x_1/z)G_{1/p}(x_2)+(x_1\leftrightarrow x_2)]\\\nn
&+&d\hat{\sigma}_{gg}^B [P_{1'1}(z,\e)G_{1/p}(x_1/z)G_{2/p}(x_2)+P_{2'2}(z,\e)G_{2/p}(x_1/z)G_{1/p}(x_2)]\}\\\nn
& \times& \frac{dz}{z}\left(\frac{1-z}{z}\right)^{-\e}dx_1 dx_2\, ,
\eeqn
where $G_{i/P}$ is the bare PDFs.

\subparagraph{Final state collinear radiation}

\begin{figure}[ht]
\begin{center}
  \includegraphics[width=1.0\textwidth]{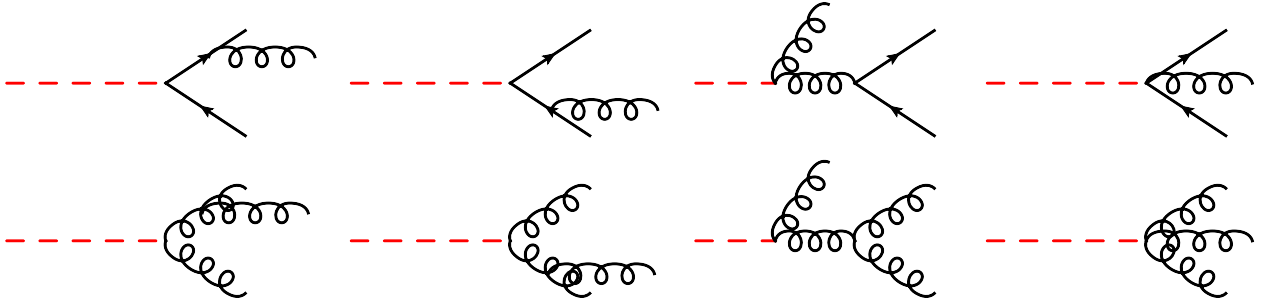}\\
  \caption{ \label{fig:fin_real} Real correction Feynman diagrams for the decay of the KK graviton.}
\end{center}
\end{figure}
The real emission diagrams from final states are shown in Fig.~\ref{fig:fin_real}. The treatment of the final state collinear singularities is much the same as that in the previous case of initial state situation. But for indistinguishable final states, there is no need to introduce fragmentation functions. For process $1+2\rightarrow 3+4+5$ with $5$ splitting from parton $4$, following similar treatment as for the initial state, we have
\be
d\sigma_{HC}^{1+2\rightarrow 3+4+5}=d\sigma_{0}^{1+2\rightarrow3+4'} \frac{\alpha_s}{2\pi}D_{\e}\left(-\frac{1}{\e}\right)\delta_c^{-\e}\int dz z^{-\e}(1-z)^{-\e}P_{44'}(z,\e)\,.
\ee
Expanding the integrand and performing the integration over $z$ yields the final state hard-collinear terms
\be
d\sigma_{HC,F}^{1+2\rightarrow 3+4+5}=d\sigma_{0}^{1+2\rightarrow3+4'} \frac{\alpha_s}{2\pi}D_{\e}\left(\frac{A_1^{4'\rightarrow 4 5}}{\e}+A_0^{4'\rightarrow 4 5}\right)\, ,
\ee
where
\beqn
A_1^{q\rightarrow qg}&=&C_F(3/2+2 \ln\delta_s)\,,\\\nn
A_0^{q\rightarrow qg}&=&C_F[7/2-\pi^2/3-\ln_{\delta_s}-\ln\delta_c(3/2+2\ln\delta_s)]\,,\\\nn
A_1^{g\rightarrow q\bar{q}}&=&-n_f /3\,,\\\nn
A_0^{g\rightarrow q\bar{q}}&=&n_f/3 (\ln\delta_c -5/3)\,,\\\nn
A_1^{g\rightarrow g g}&=&C_A(11/6+2\ln\delta_s)\,,\\\nn
A_0^{g\rightarrow g g}&=&C_A[67/18-\pi^2/3-\ln^2\delta_s-\ln\delta_c(11/6+2\ln\delta_s)]\,.
\eeqn

\paragraph{\normalsize{\textbf{Hard non-collinear emission}}}
We also have to consider contributions from the hard non-collinear part, which is finite. The hard non-collinear partonic cross section is given by
\begin{equation}
\label{eq:cocr}
\hat{\sigma}_{ij}^{\overline{\rm HC}}\,=\,\frac{1}{2\,s}\int_{\overline{\rm HC}} \overline{\sum}\,|M_{ij}^3|^2\,dPS_3\,.
\end{equation}
We can calculate the amplitude square of these real radiation diagrams directly in 4 dimensions. Besides the channels we have considered in the LO order diagram, there are also $q\,g$ and $\bar{q}\,g$ initial state processes. The detail results are given in the appendix.

\subsubsection{Mass Factorization}

After adding the renormalized virtual corrections and the two-cut off real corrections, the parton level cross section still contain some collinear divergences which can be absorbed into a redefinition of the PDFs at the NLO, namely mass factorization~\cite{altarelli}. This procedure means we replace the bare PDF $G_{a/p}(x)$ with renormalized PDF $G_{a/p}(x,\mu_f)$ and then convolute it with the partonic cross section. With the $\overline{MS}$ convention the scale-dependent PDF $G_{a/p}(x,\mu_f)$ is given by~\cite{Harris:2001sx}
\be
G_{a/p}(x,\mu_f)=G_{a/p}(x)+\sum_b\left(\frac{1}{\e}\right)\left[\frac{\alpha_s}{2\pi}\frac{\Gamma(1-\e)}{\Gamma(1-2\e)}\left(\frac{4\pi\mu_r^2}{\mu_f^2}\right)^\e\right]\int_x^1\frac{dz}{z}P_{ab}(z)G_{b/p}(x/z)
\ee

This replacement will produce a collinear singular term, which will be combined with the hard collinear contribution in Eq.~(\ref{eq:hc}). Then the expression for the remaining collinear contribution after considering $gg$ initial state contribution will be:
\beqn
&&d\sigma_{ij}^{\rm coll,I}= \frac{\alpha_s}{2\pi}D_{\e}\{[\tilde{G}_{q/p}(x_1,\mu_f)G_{\bar{q}/p}(x_2,\mu_f)+G_{q/p}(x_1,\mu_f)\tilde{G}_{\bar{q}/p}(x_2,\mu_f)\\\nn
&&\,+\sum_{\alpha=q,\bar{q}}\left[\frac{A_1^{sc}(\alpha\rightarrow \alpha g)}{\e}+A_0^{sc}(\alpha\rightarrow\alpha g)\right]G_{q/p}(x_1,\mu_f)G_{\bar{q}/p}(x_2,\mu_f)+(x_1\leftrightarrow x_2)]d\hat{\sigma}_{q \bar{q}}^B \\\nn
&&\,+[\tilde{G}_{g/p}(x_1,\mu_f)G_{g/p}(x_2,\mu_f)+G_{g/p}(x_1,\mu_f)\tilde{G}_{g/p}(x_2,\mu_f)\\\nn
&&\,+2\left[\frac{A_1^{sc}(g\rightarrow g g)}{\e}+A_0^{sc}(g\rightarrow g g)\right]G_{g/p}(x_1,\mu_f)G_{g/p}(x_q,\mu_f)]d\hat{\sigma}_{gg}^B \}dx_1 dx_2\,,
\eeqn
where
\beqn
A_1^{sc}(q\rightarrow qg)&=&A_1^{sc}(\bar{q}\rightarrow \bar{q}g)=C_F(3/2+2 \ln\delta_s)\, ,\\\nn
A_1^{sc}(g\rightarrow gg)&=&2 C_A \ln\delta_s+(11C_A-2n_f)/6\, ,\\\nn
A_0^{sc}&=&A_1^{sc}\ln\left(\frac{s}{\mu_f^2}\right)\, ,\\\nn
\tilde{G}_{a/p}(x,\mu_f)&=&\sum_{a'}\int_{x}^{1-\delta_s \delta_{a a'}} \frac{d y}{y} G_{a'/p}(x/y,\mu_f)\tilde{P}_{a a'}(y)\, ,\\\nn
\widetilde{P}_{i j}(y)&=&P_{i j}(y)\ln(\delta_c \frac{1-y}{y}\frac{s}{\mu_f^2})-P'_{i j}(y)\, .
\eeqn

Finally, the NLO total cross section for $p\,p\rightarrow j\,j$ in the $\overline{MS}$ factorization scheme is,
\beqn
\sigma^{NLO}&=&\int dx_1dx_2[G_{q/p}(x_1,\mu_f)G_{\bar{q}/p}(x_2,\mu_f)+(x_1\leftrightarrow x_2)](\hat{\sigma}_{q\bar{q}}^B+\hat{\sigma}_{q\bar{q}}^V+\hat{\sigma}_{q\bar{q}}^S+\hat{\sigma}_{q\bar{q}}^{HC,F}\\\nn
&+&\hat{\sigma}_{q\bar{q}}^{\overline{HC}})+\int dx_1dx_2 G_{g/p}(x_1,\mu_f)G_{g/p}(x_2,\mu_f)(\hat{\sigma}_{gg}^B+\hat{\sigma}_{gg}^V+\hat{\sigma}_{gg}^S+\hat{\sigma}_{gg}^{HC,F}+\hat{\sigma}_{gg}^{\overline{HC}})+\hat{\sigma}^{coll,I}\\\nn
&+&\int dx_1dx_2 \sum_{\alpha=q,\bar{q}}[G_{g/p}(x_1,\mu_f)G_{\alpha/p}(x_2,\mu_f)+(x_1\leftrightarrow x_2)]\hat{\sigma}_{\alpha g}^{\overline{HC}})
\eeqn

Note that the above expression contains no singularities since $2 A_2^v+A_2^s=0$ ,$2A_1^{v,q}+A_1^{s,q}+2A_1^{sc}(q\rightarrow qg)=0$, $2A_1^{v,g}+A_1^{s,g}+2A_1^{sc}(g\rightarrow gg)=0$ for initial state calculation. And similar results can be obtained for final states.

\subsubsection{Consistent treatment of KK graviton decay in perturbation theory}\label{s4}

In the narrow width approximation (NWA)~\cite{Cao:2004ap}, the production cross section for a specific decay channel is given by the total cross section times the branching fraction of the decay channel, which requires a consistent treatment of the decay at the NLO. For the Breit-Wigner resonance there is a similar procedure. In this subsection, we briefly review the basic idea of this procedure in NWA, then introduce the method we use for Breit-Wigner resonance.

The perturbative expansion of cross section and decay width can be written as,
\beqn
\sigma^{NLO}=\sigma_0+\alpha_s \sigma_{1}\\\nn
\Gamma^{NLO}=\Gamma_0+\alpha_s\Gamma_1\\
\eeqn

Following the approach in Ref.~\cite{Cao:2004ap}, by expanding the cross section to $\mathcal{O}(\alpha_s)$ and discarding terms of order $\mathcal{O}(\alpha_s^2)$ or higher, we can write the differential cross sections as,
\be
\sigma_i^{NLO}=\sigma_0\times\frac{\Gamma_0^i}{\Gamma_0}+\sigma_0\times\frac{\alpha_s \Gamma_1^i}{\Gamma_0}+\alpha_s\sigma_1\times\frac{\Gamma_0^i}{\Gamma_0}-\alpha_s\sigma_0\times\frac{\Gamma_0^i}{\Gamma_0}\frac{\Gamma_1}{\Gamma_0},
\ee
where $\sigma_0$ and $\Gamma_0$ are the lowest order contributions to the production rate and total decay width and $\alpha_s\sigma_1$ and $\alpha_s\Gamma_1$ the corresponding NLO corrections. Meanwhile, $\Gamma^{i}_0$ and $\alpha_s$$\Gamma^{i}_1$ is the LO differential decay width and its NLO corrections for the channel $i$ we considered.
Following the above approach, we expand the KK graviton propagator with NLO decay width as
\beqn
&&\frac{1}{(s-m_{KK}^2)^2+[\Gamma_0+\alpha_s\Gamma_1]^2 m_{KK}^2}\\\nn
&=&\frac{1}{(s-m_{KK}^2)^2+\Gamma_0^2 m_{KK}^2}-\frac{2 \alpha_s m_{KK}^2 \Gamma_0 \Gamma_1}{[(s-m_{KK}^2)^2+\Gamma_0^2 m_{KK}^2]^2}\,,\\\nn
&=&R(s)\left[1-2 \alpha_s R(s) m_{KK}^2 \Gamma_0 \Gamma_1\right]\,,
\eeqn
then we can rewrite similar cross section for Breit-Wigner resonance as
\be
  \sigma_{i}^{NLO} =   \left[1-2 \alpha_s R(s) m_{KK}^2 \Gamma_0 \Gamma_1\right]\sigma^{0}\otimes\Gamma^{i}_0 +  \alpha_s \sigma^{1}\otimes\Gamma^{i}_0+\alpha_s \sigma^{0}\otimes\Gamma^{i}_1,
\ee
where in the convolution the LO width is always used in the propagator.

\begin{figure}[ht]
\begin{center}
  \includegraphics[width=1.0\textwidth]{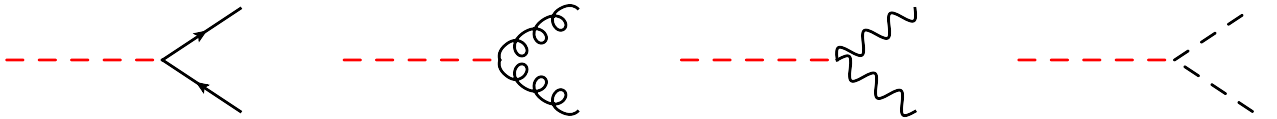}\\
  \caption{ \label{width} Decay channels of KK graviton.}
\end{center}
\end{figure}

\begin{figure}[ht]
\begin{center}
  \includegraphics[width=1.0\textwidth]{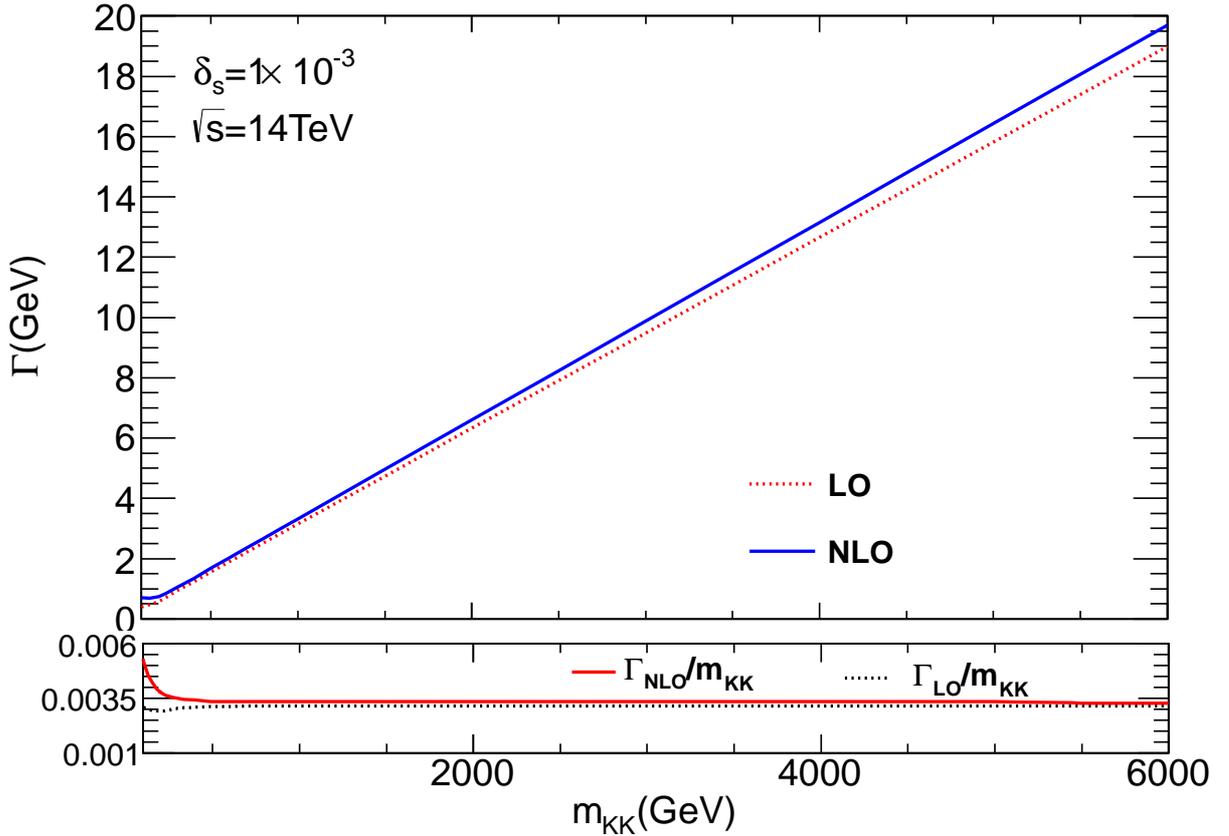}\\
  \caption{ \label{width1} The LO and NLO Decay width of the KK graviton for different KK graviton mass.}
\end{center}
\end{figure}

Now we turn to the calculations of NLO QCD corrections for the decay width of KK graviton. The  KK graviton can decay to all the particles in SM, which is shown in Fig.~\ref{width}. The LO decay width has been calculated in Ref~{\cite{Randall:1999ee}}, and the calculation of NLO total decay width is straight forward. Fig.~\ref{width1} shows the mass dependence of the LO and NLO decay width, which can be fitted as,
\beqn\label{eq:width}
\Gamma_0&=&3.15\times10^{-3}\,*\,m_{KK}\,,\\\nn
\alpha_s\Gamma_1&=&2.08\times10^{-3}\,*\,\alpha_s\,m_{KK}\,.
\eeqn

\section{Numerical Results}\label{s5}

\subsection{Cross section}
In this subsection, we present the numerical results for total and differential cross sections for dijet production via RS KK graviton at the LHC. In our numerical calculations, we use the two-loop evaluation for $\alpha_s(Q)$ ~\cite{Gorishny:1990zu} and CTEQ PDFs~\cite{Pumplin:2002vw}. We use CTEQ6M PDF for NLO calculation and CTEQ6L PDF for LO calculation in our numerical calculations of total and differential cross sections, respectively. We assume $k/{\bar{M}_{p}}=0.1$ and $m_{KK}=1.5\, \rm {TeV}$ or $2 \,\rm {TeV} $ for the RS model unless specified, so the coupling strength between the graviton and the Standard Model particles will be fixed when the graviton mass is set, as shown in Eq.~(\ref{l m x1}).

For the final-state jets, we use the anti-$k_t$ jet algorithm~\cite{Cacciari:2008gp} with the distance parameter $D=0.5$ to combine QCD partons into jets. We reconstruct the trigger jet using the FASTJET algorithm~\cite{Cacciari:2011ma}. We also require the final-state jets to satisfy the following basic kinematic cuts according to ones used in the CMS study~\cite{Chatrchyan:2013qha}
\begin{eqnarray}
&&p_{T_{j}}>30 {\rm~GeV},\quad|\eta_j|<2.5.\nonumber
\end{eqnarray}
Here $p_{T_{j}}$ and $\eta_{j}$ are the transverse momentum and pseudorapidity of the final state jets, respectively.

Both the renormalization and factorization scales are fixed to the invariant mass $m_{jj}$ of the dijet final states, where $m_{jj}=\sqrt{(E_{j1}+E_{j2})^2-|\overrightarrow{p}_{j1}+\overrightarrow{p}_{j2}|^2}$.

We have checked that the Breit-Wigner approximation is applicable at the LO, and calculated the full LO results including all the channels. The results show that the contribution from the $s$ channel, which we discuss in our work, is dominant, since contributions from $t$ and $u$ channel are about $6\%$ of full LO total cross section. After taking experiment kinematics requirement shown below, their contributions are extremely smaller, which are only about $3\%$. The reason is that $t$ and $u$ channels are obviously suppressed by the kinematics effect from the KK graviton propagator.

\begin{figure}[ht]
\begin{center}
  \includegraphics[width=1.0\textwidth]{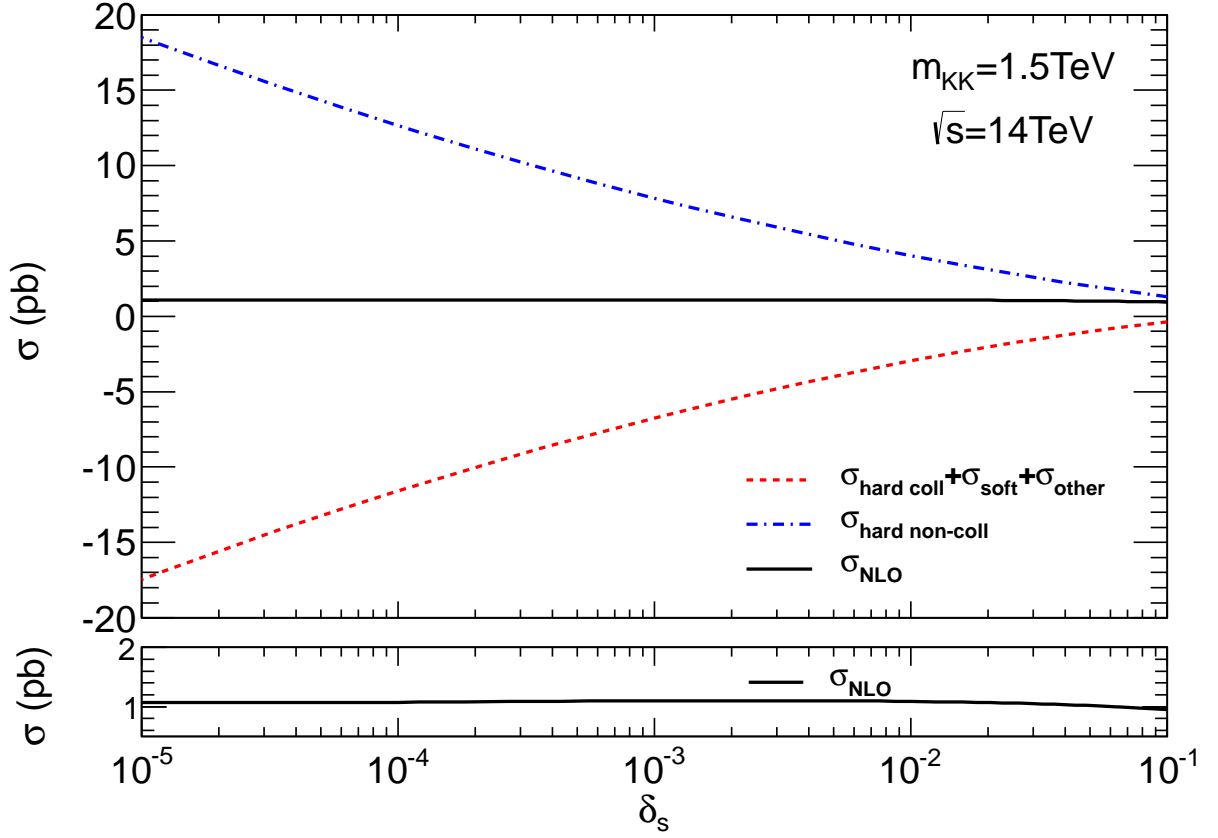}\\
  \caption{ \label{deltas} Total cross sections for $p\,p\rightarrow G \rightarrow j\,j$ at the LHC as a function of $\delta_s$ in the phase space slicing treatment. The $\delta_c$ is chosen to be $\delta_c=\delta_s/50$.}
\end{center}
\end{figure}
In  Fig.~\ref{deltas} we show that the dependence of NLO total cross section on the arbitrary cutoff $\delta_s$ and $\delta_c$ is indeed very weak. Here $\sigma_{other}$ includes the contribution from the Born cross section and the virtual corrections. Both the soft plus hard collinear contributions and the hard non-collinear contributions depend strongly on the cutoffs and, especially for the small cutoffs ($\delta_s<10^{-2}$). However, after combining every contribution ($\sigma_{\rm soft}+\sigma_{\rm hard-coll}+\sigma_{\rm virtual}+\sigma_{\rm hard-noncoll}$), such dependence on the cutoffs cancel each other. The final results for $\sigma_{NLO}$ are almost independent of the cutoff for $\delta_s<10^{-2}$. We take $\delta_s=10^{-3}$ and $\delta_c=\delta_s/50$ to obtain the numerical results presented below.

\begin{figure}[ht]
\begin{centering}
\includegraphics[width=0.45\textwidth]{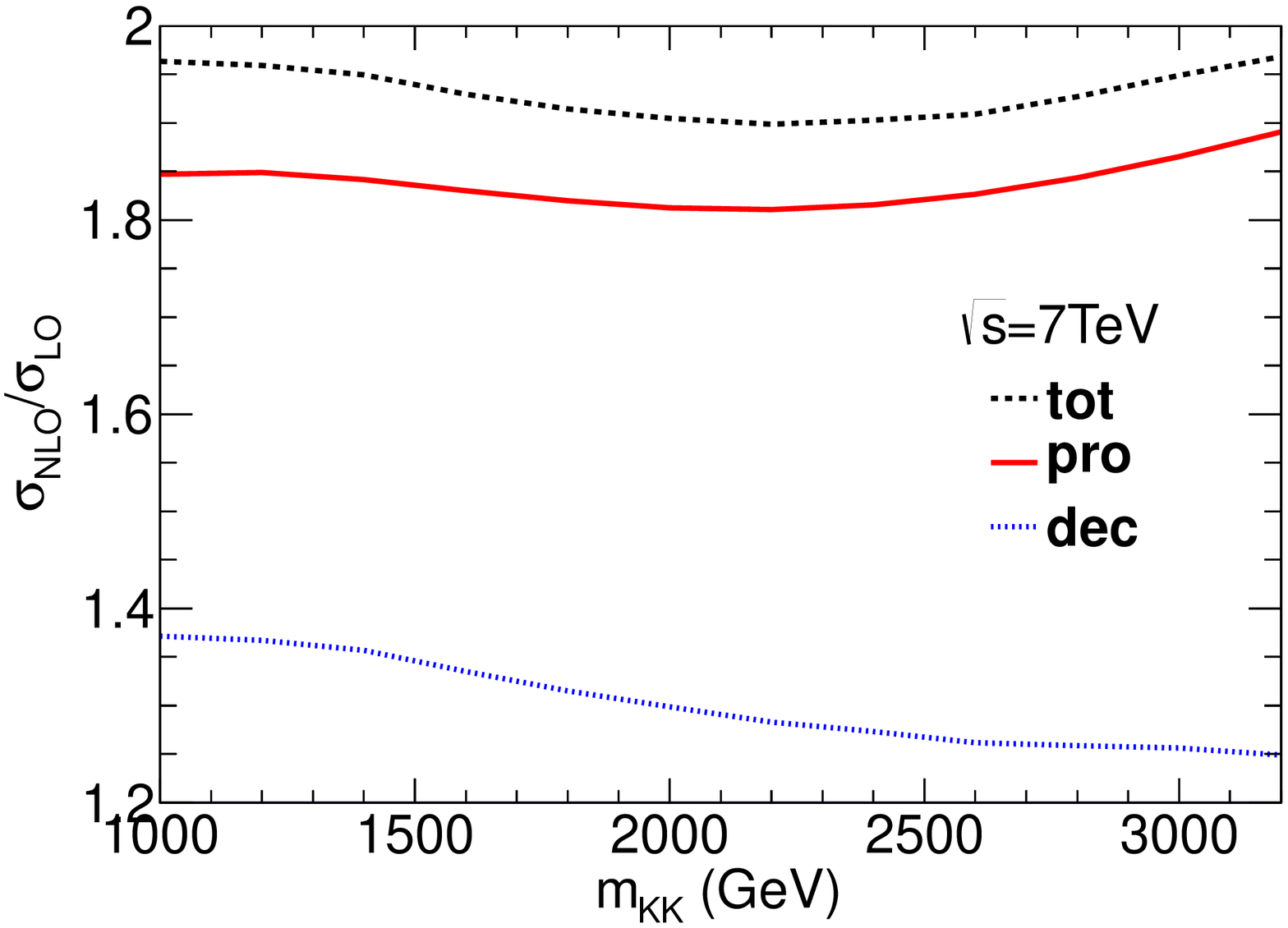}~~\includegraphics[width=0.45\textwidth]{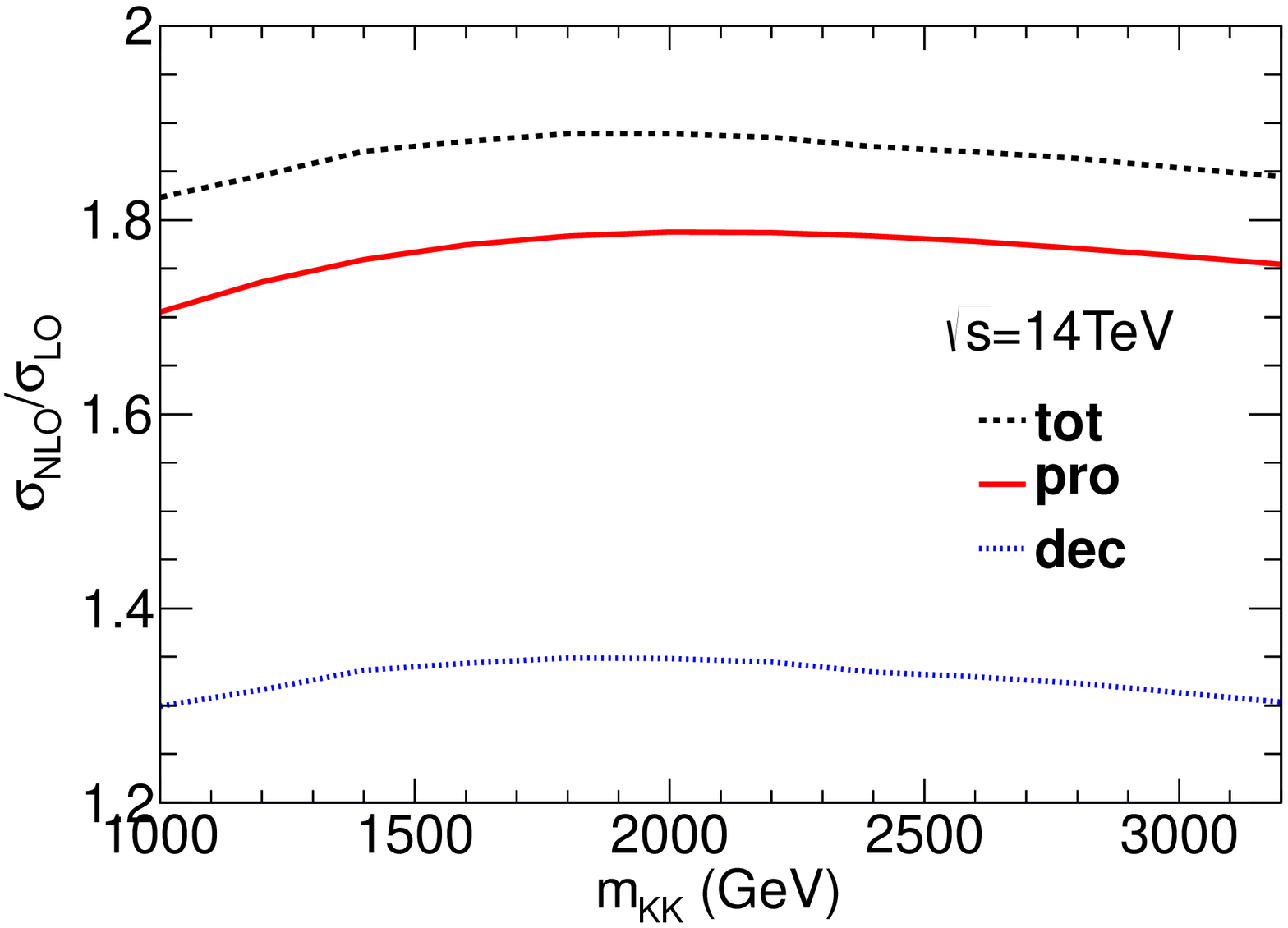}
\par\end{centering}

\vspace{-1ex}
 \caption{\label{fig:kfactor}The NLO K factors as functions of the heavy resonance mass at the LHC. The long dotted and solid lines correspond to including the total NLO QCD corrections and the corrections from the production part alone, respectively. }
\end{figure}
Fig.~\ref{fig:kfactor} shows the NLO K factor, which defined as the ratio of the NLO cross section $\sigma_{NLO}$ to the LO cross section $\sigma_{LO}$, as a function of the KK graviton mass at the LHC with different center of mass energies. We can see that the total QCD NLO corrections can be large, which can increase the total cross sections by about $80\%-100\%$. Numerical results show that the NLO corrections from the production part are dominant, and agree with the ones given in Refs.~\cite{Li:2006yv,Agarwal:2010sn,Gao:2010bb}. The contributions from the decay part are relatively small, but can still reach about $20\%-30\%$.

\begin{figure}[ht]
\begin{centering}
\includegraphics[width=0.45\textwidth]{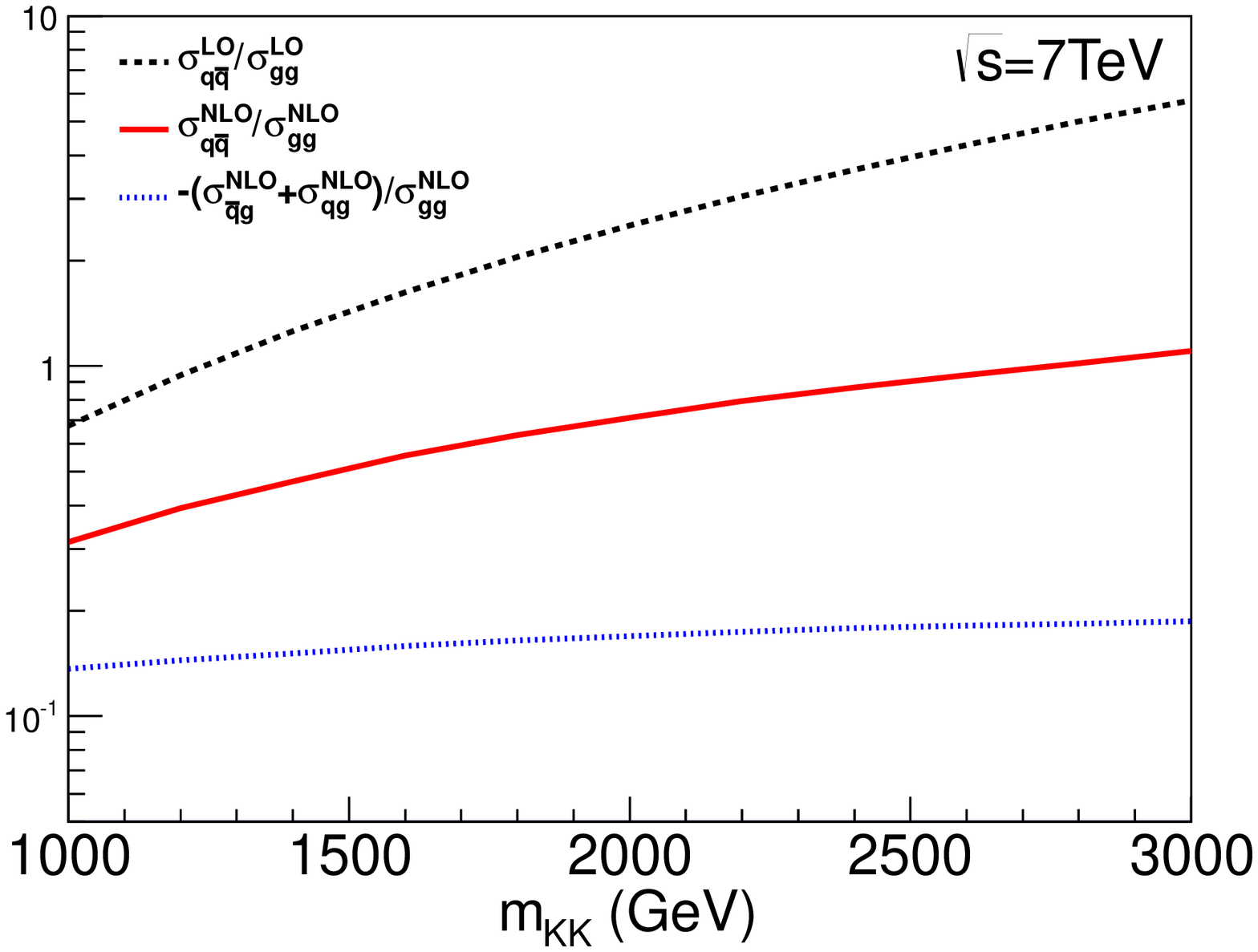}~~\includegraphics[width=0.45\textwidth]{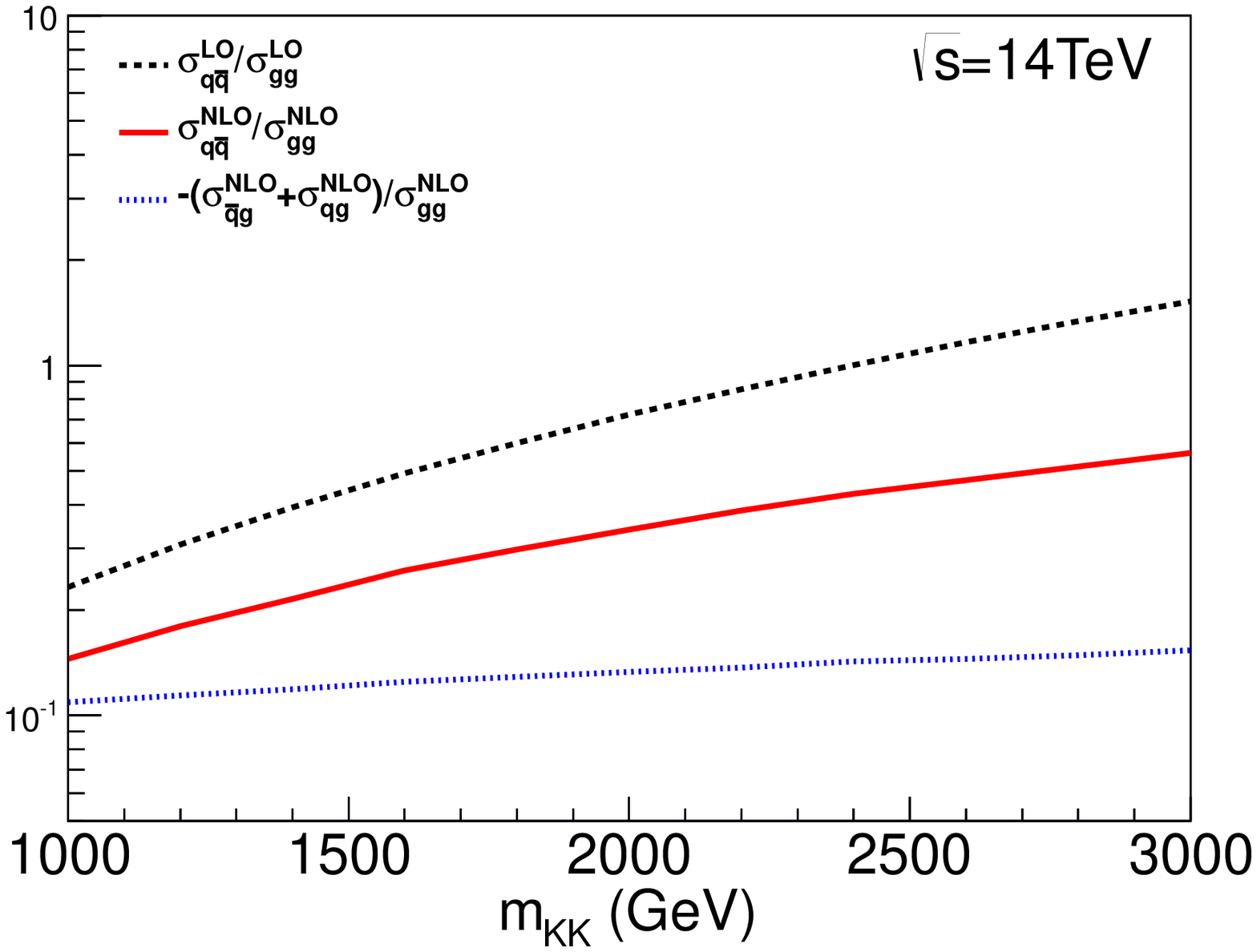}
\par\end{centering}

\vspace{-1ex}
 \caption{\label{fig:chan}The ratios of the total cross sections from different channels for the graviton as functions of the graviton mass at both the LO and the NLO. }
\end{figure}

We further present the ratios between the total cross sections from the different channels at both the LO and the NLO in Fig.~\ref{fig:chan} . It can be found that the contribution from the $gg$ channel is dominant at low KK graviton mass region for the large PDF of the gluon, and the contribution from the $q\bar{q}$ channel becomes more important at the high mass region since the PDF of the valence quark decreases more slowly than the gluon. The NLO corrections can change the ratio between different channels significantly.

\begin{figure}[ht]
\begin{centering}
\includegraphics[width=0.45\textwidth]{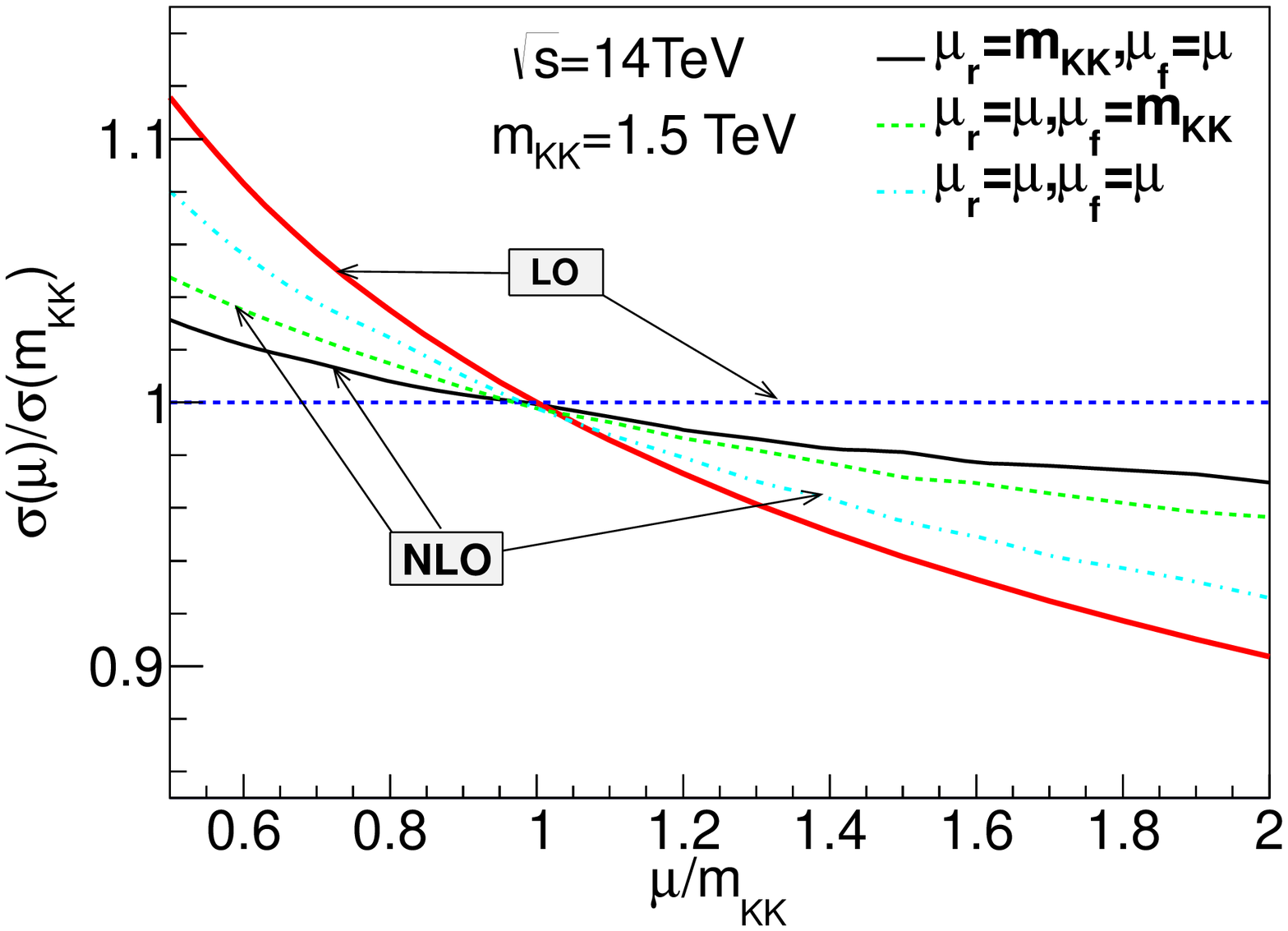}~~\includegraphics[width=0.45\textwidth]{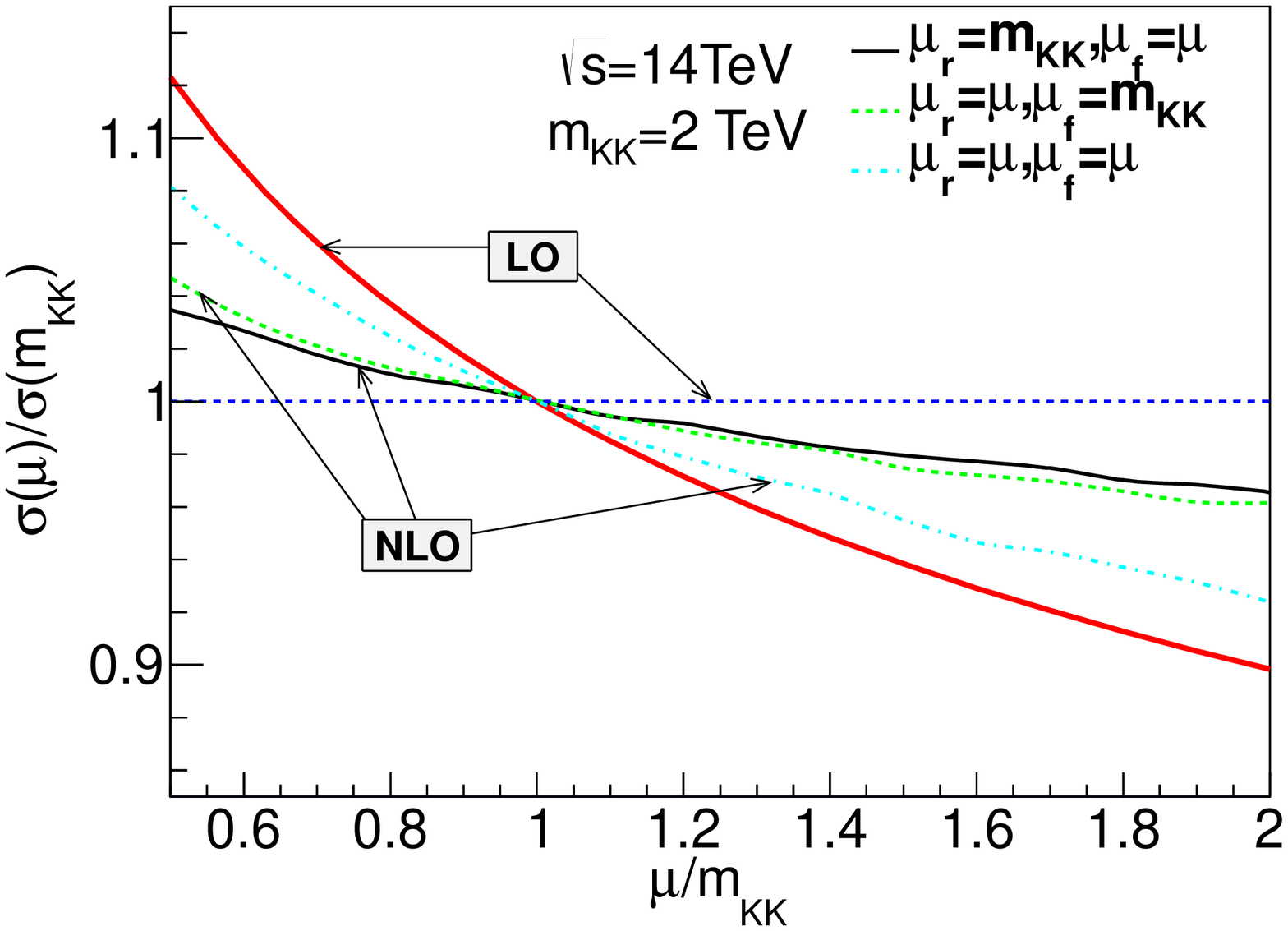}
\par\end{centering}

\vspace{-1ex}
 \caption{\label{fig:scale} Scale dependence of the total cross sections for dijet production through RS KK graviton at the LHC with different KK graviton mass.}
\end{figure}

In Fig.~\ref{fig:scale}, we show scale dependencies of the LO and NLO total cross sections. At the LO, the scale dependence is purely from the factorization scale. Fig.~\ref{fig:scale} shows that the factorization scale dependence of the NLO cross section is significantly reduced compared with the LO result.
\subsection{Differential cross section}
We separately present invariant mass and transverse momentum distribution in this subsection. Following the experimental analysis in~\cite{Chatrchyan:2013qha}, we consider wide jets as the final states, which are formed by clustering additional jets into the closest leading jet if within a distance $\Delta R = \sqrt{\Delta \eta^2 + \Delta \phi^2}< 1.1$. To account for resolution of the detectors, we also add a Gaussian smearing to the energy of final-state jets~\cite{Aad:2009wy}, where the width is set as
\begin{eqnarray}\label{E_smear}
&&\Delta E_{j}/E_{j} = 0.5/\sqrt{E_{j}/{\rm GeV}}\oplus0.02\, .
\end{eqnarray}

\begin{figure}[ht]
\begin{centering}
\includegraphics[width=0.45\textwidth]{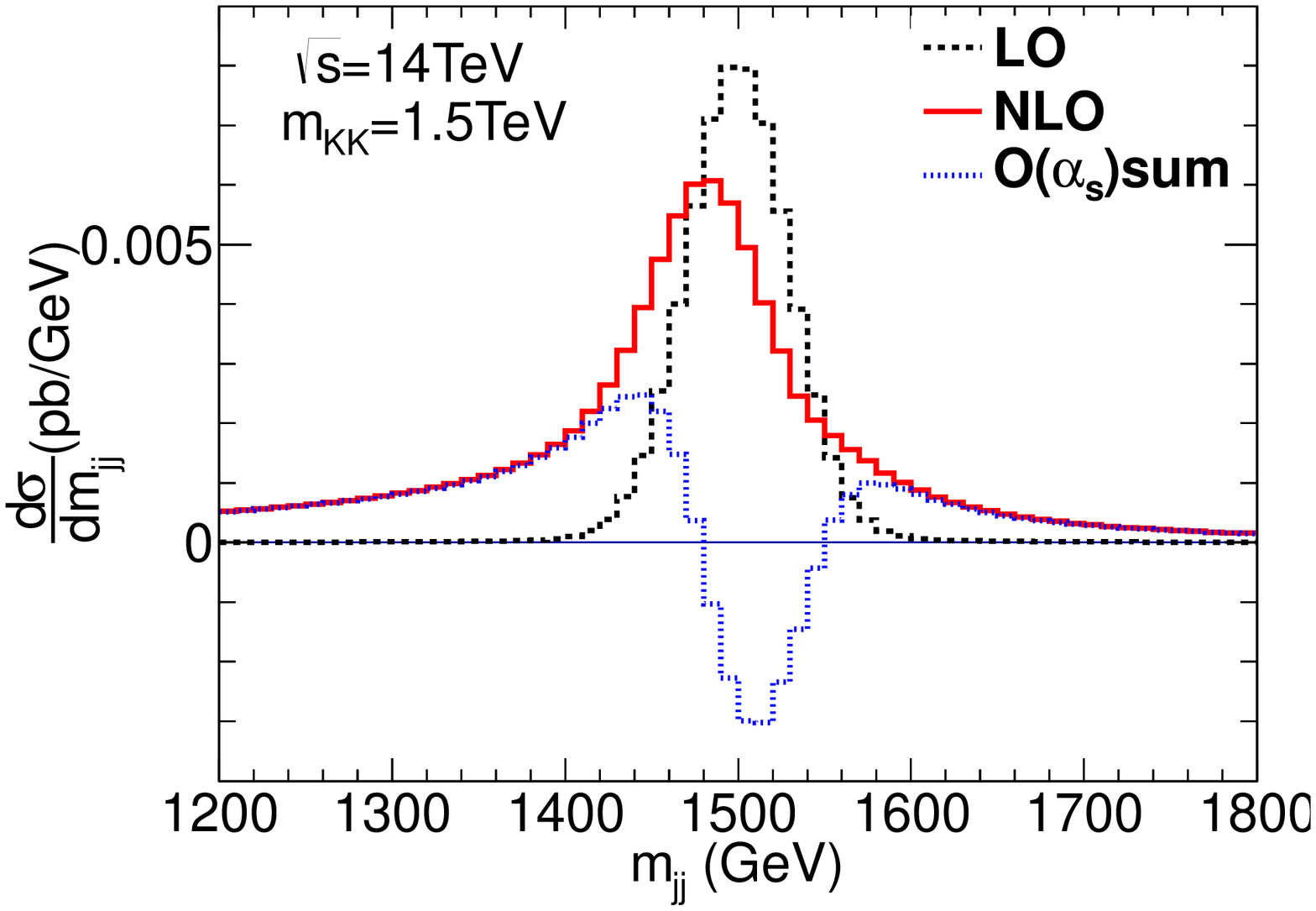}~~\includegraphics[width=0.45\textwidth]{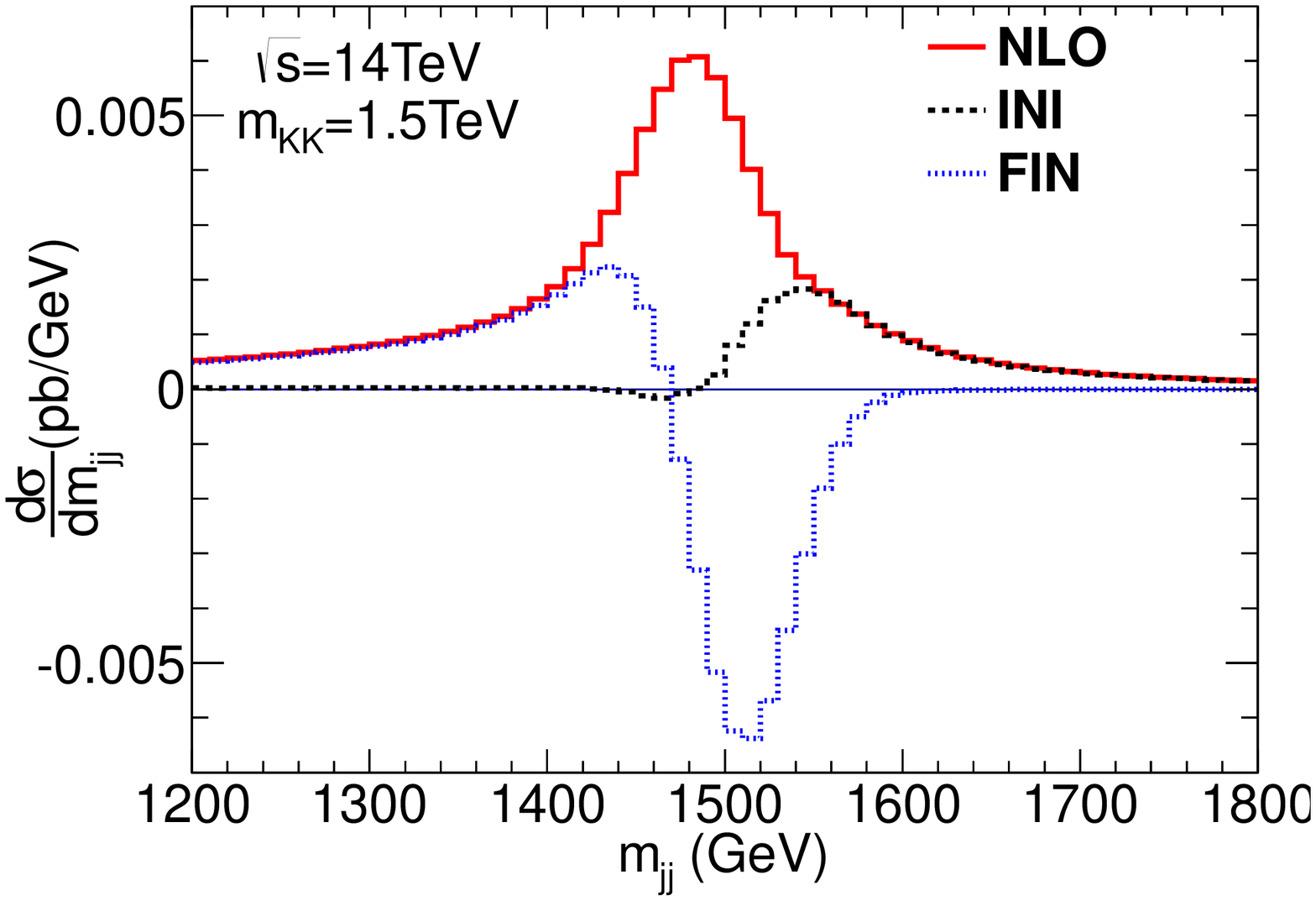}
\par\end{centering}
\vspace{-1ex}
 \caption{\label{fig:mjco}Differential cross sections in the invariant mass for the final state dijet through RS KK graviton, left plot shows the LO, NLO results and the NLO corrections. The right plot shows the NLO corrections from production and decay separately.}
\end{figure}

\begin{figure}[ht]
\begin{centering}
\includegraphics[width=0.45\textwidth]{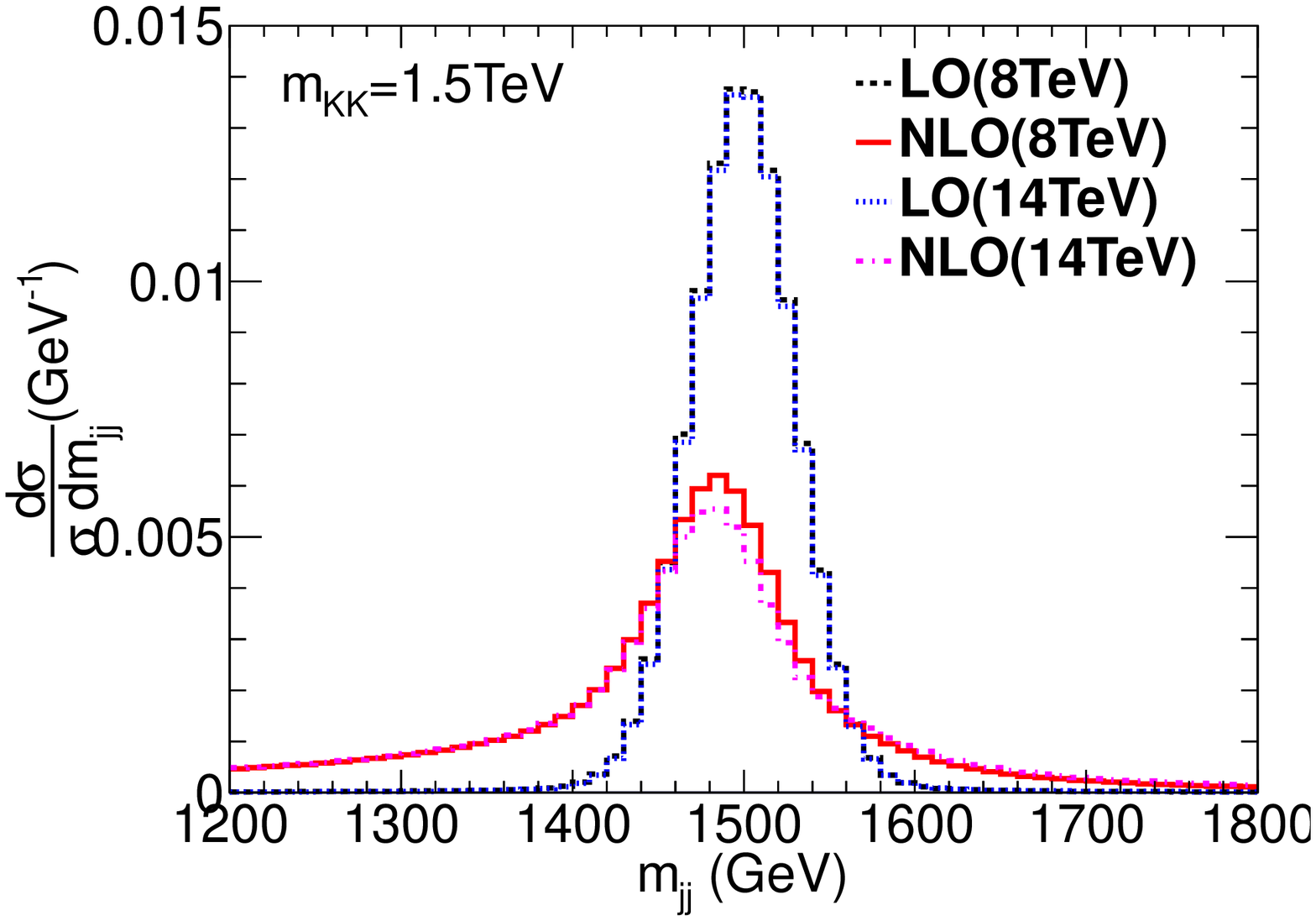}~~\includegraphics[width=0.45\textwidth]{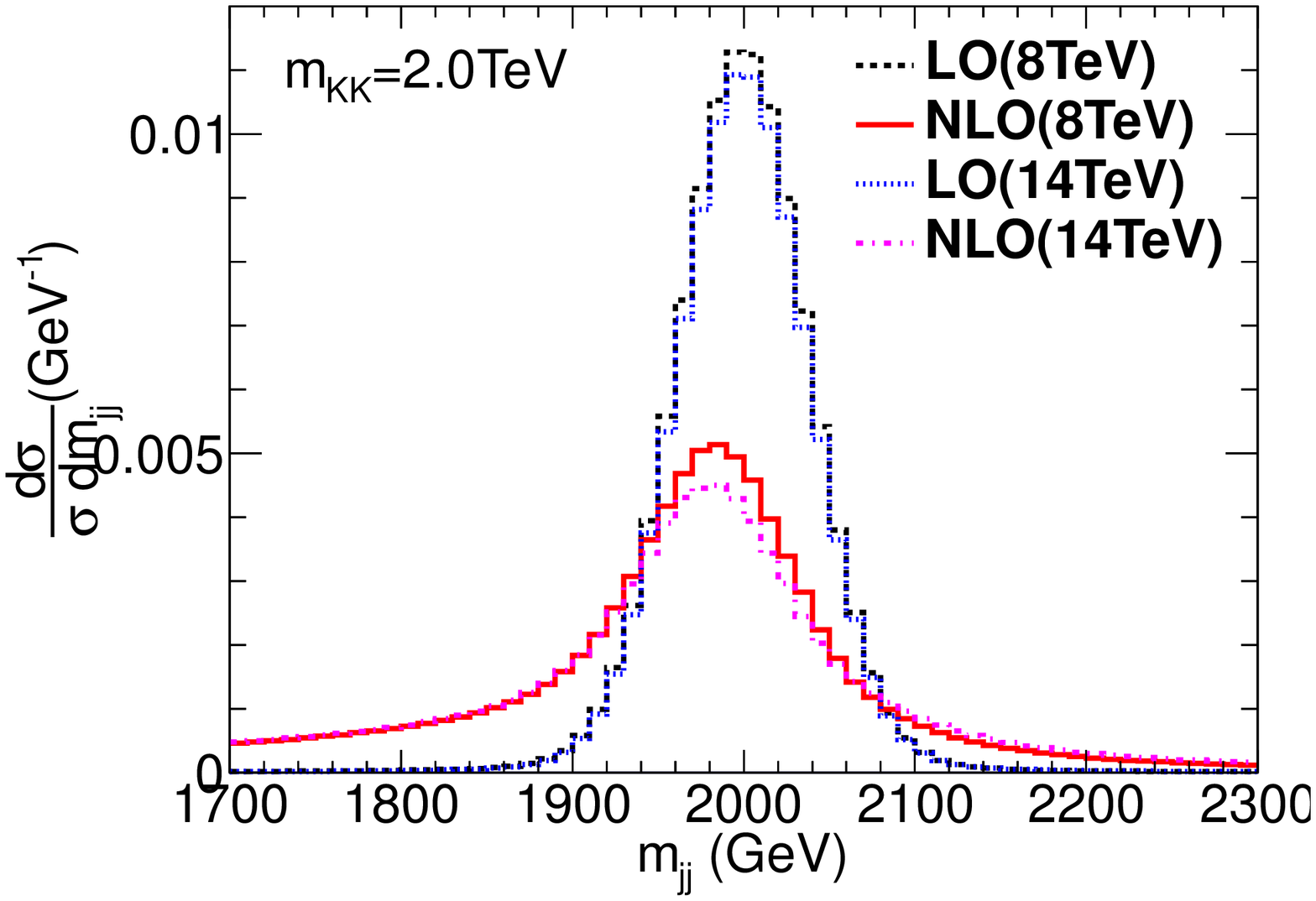}
\par\end{centering}
\vspace{-1ex}
 \caption{\label{fig:mjdm}Normalized differential cross sections of the invariant mass for the final state dijet through RS KK graviton with $m_{KK}=1.5\rm TeV$ and $2\rm TeV$.}
\end{figure}

 Fig.~\ref{fig:mjco} gives the invariant mass distributions of the dijet. At the LO it is a Breit-Wigner distribution with a center value $m_{KK}$ and width $\Gamma_{KK}$. At the NLO there could exist an additional hard parton besides the two leading jets in the final state. Thus the NLO corrections push the peak of the distributions to the lower invariant mass region.
We also show separate contributions from initial state and final state corrections in Fig.~\ref{fig:mjco}. It can be seen that the initial state corrections shift the invariant mass distributions to higher region while the final state corrections tend to shift it in opposite way, which is a consequence of different origins of the additional radiated parton.

Fig.~\ref{fig:mjdm} shows the normalized invariant mass distributions with different KK graviton mass and collider energy. Collider energy shows weak impact on the shape of the distribution.

\begin{figure}[ht!]
\begin{minipage}[t]{0.45\linewidth}
\centering
  \includegraphics[width=1.0\linewidth]{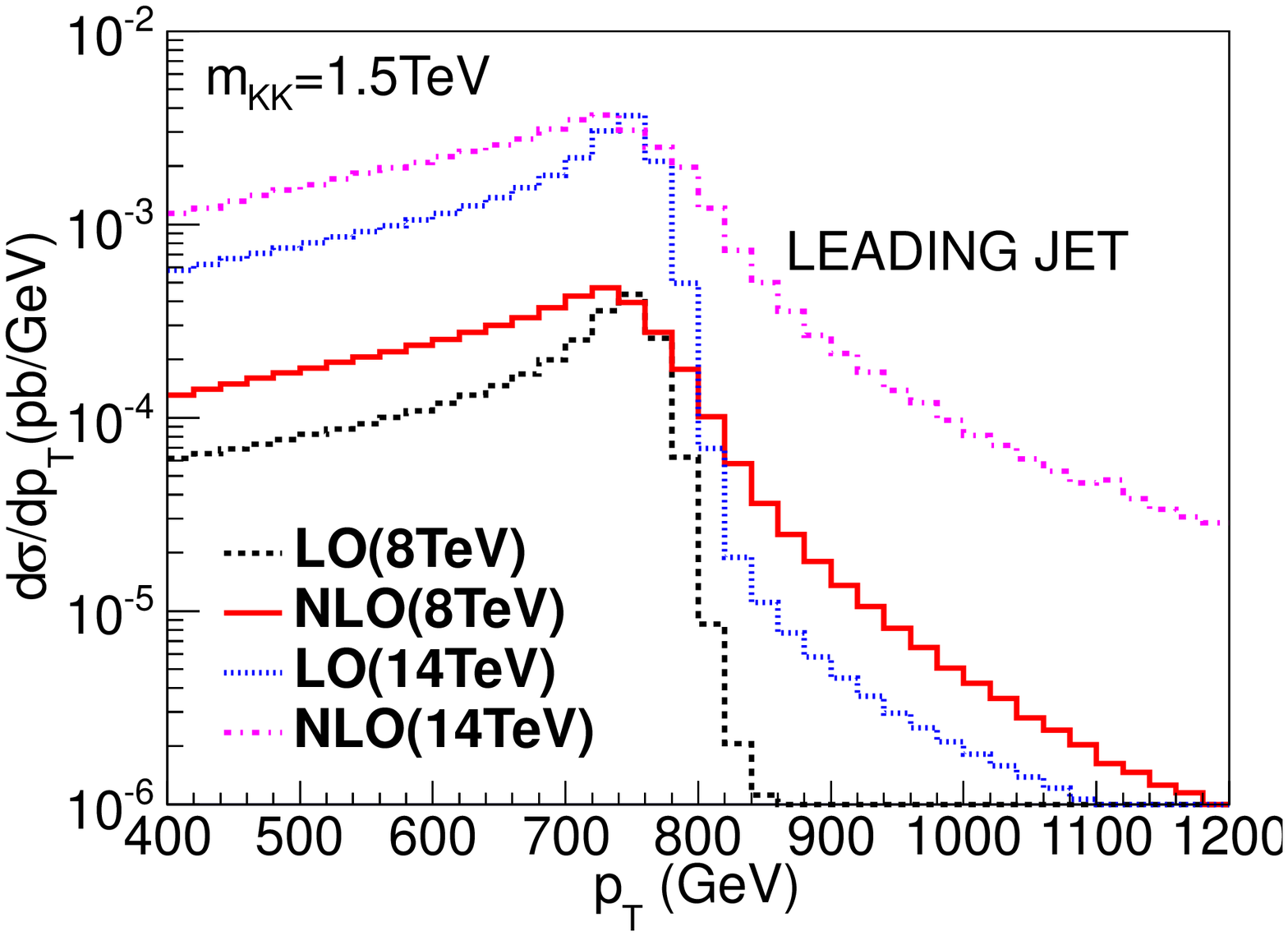}\\
\end{minipage}
\hfill
\begin{minipage}[t]{0.45\linewidth}
\centering
 \includegraphics[width=1.0\linewidth]{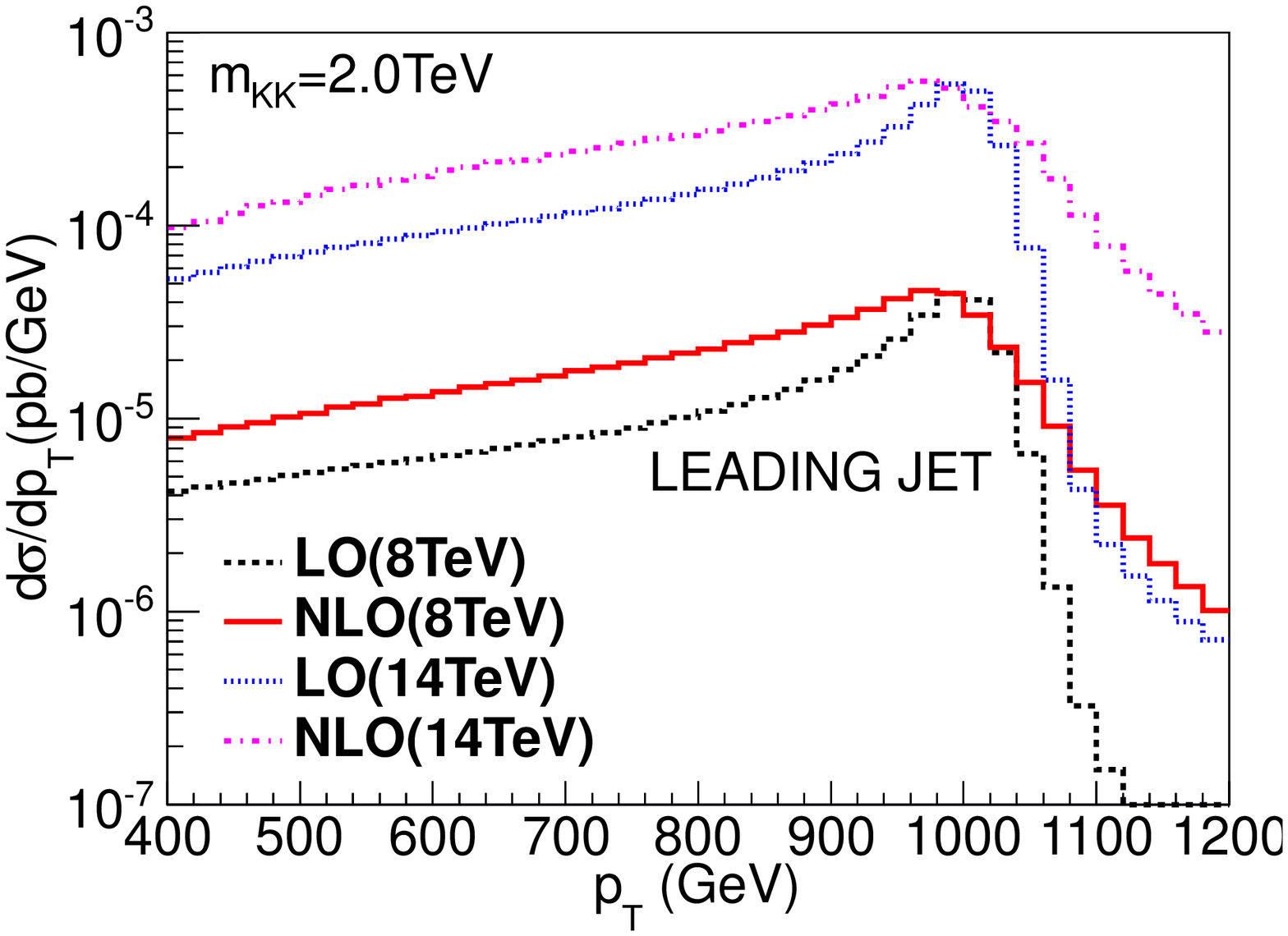}\\
\end{minipage}
\hfill
\begin{minipage}[t]{0.45\linewidth}
\centering
 \includegraphics[width=1.0\linewidth]{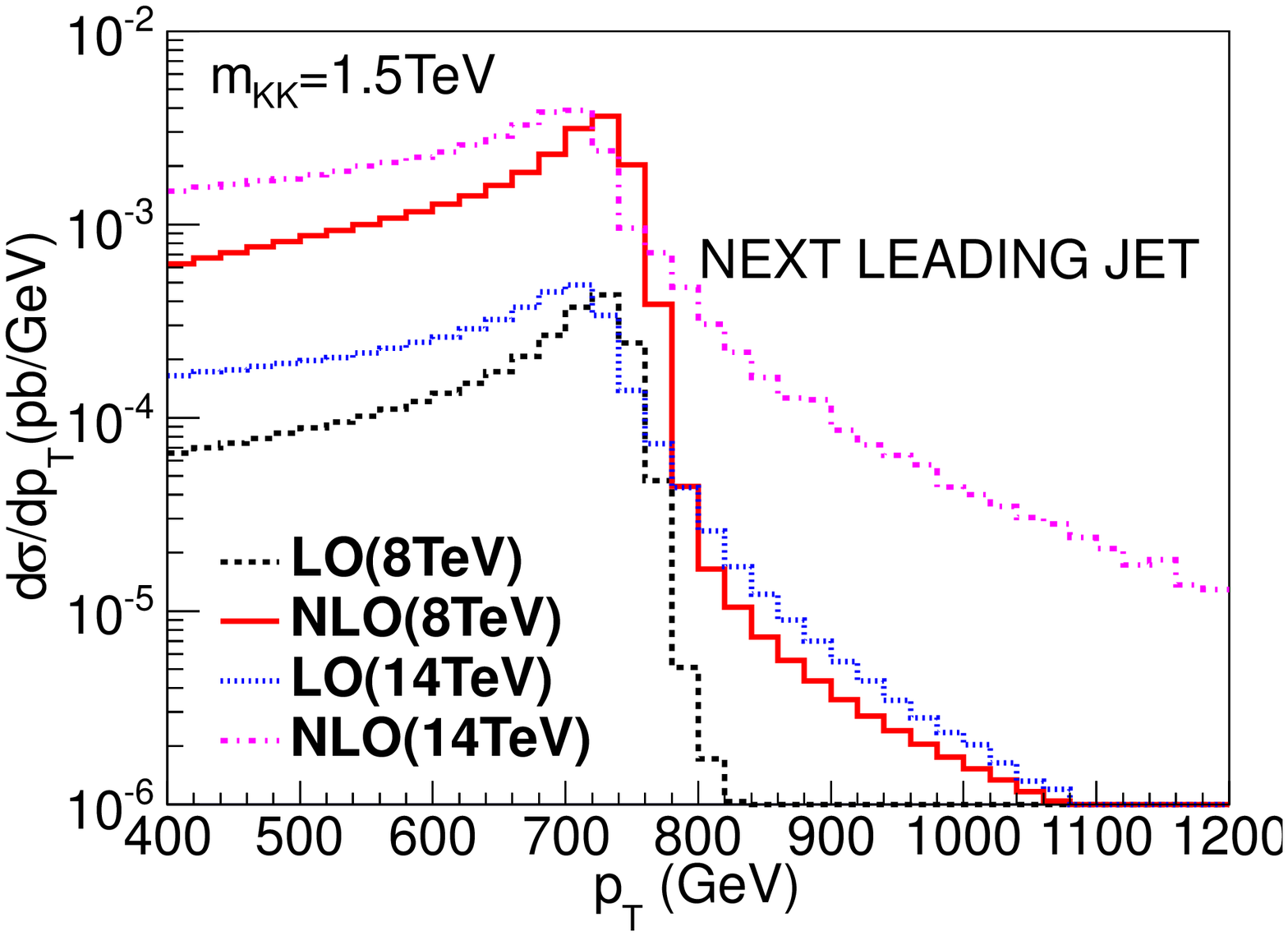}\\
\end{minipage}
\hfill
\begin{minipage}[t]{0.45\linewidth}
\centering
 \includegraphics[width=1.0\linewidth]{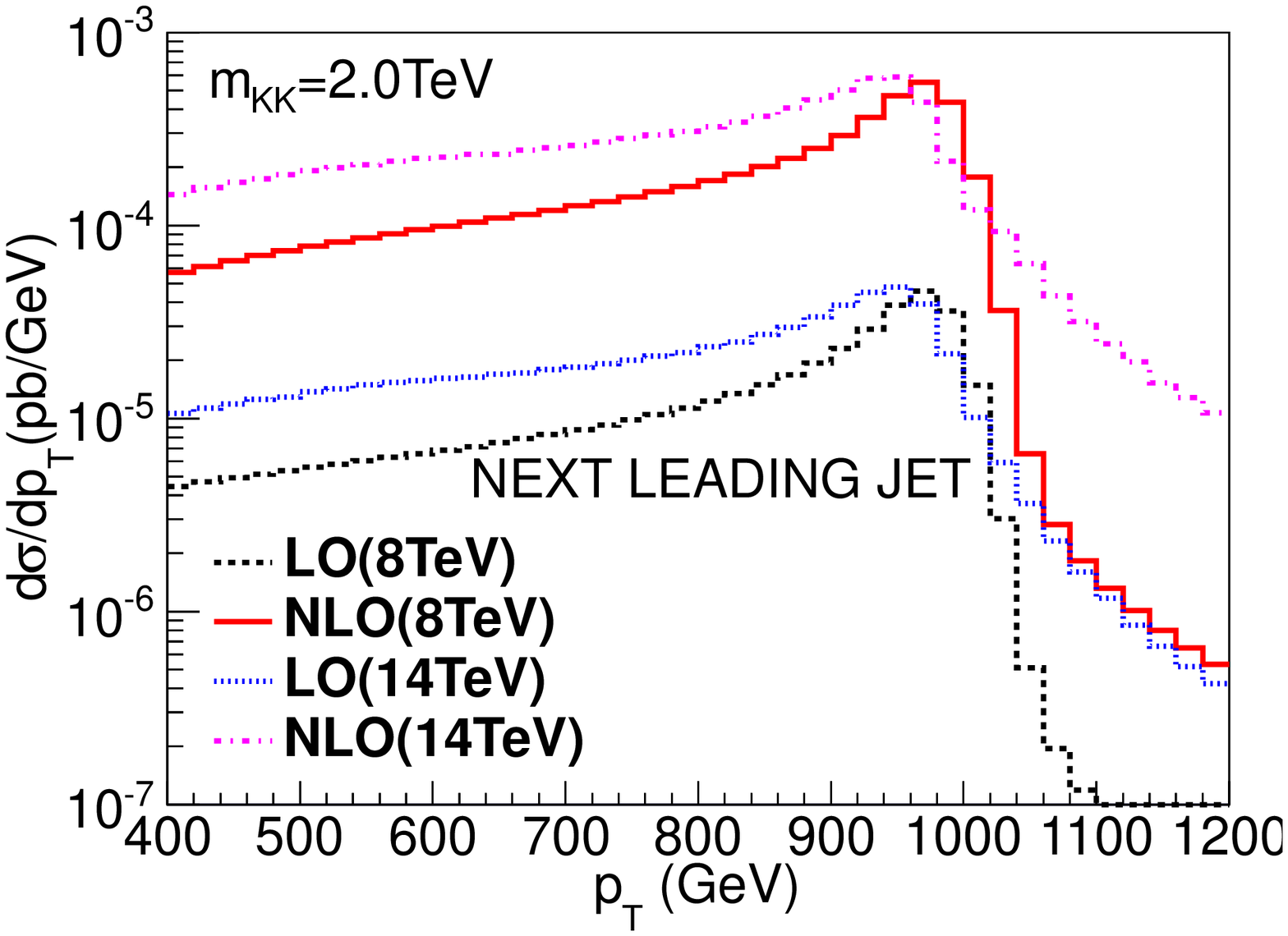}\\
\end{minipage}
\caption{\label{fig:dispe}The LO and NLO differential cross sections in the transverse momentum $p_T$ of the two leading jets in dijet production. Different type of line corresponds to different collider energy. The top row shows the result for the leading jet while the bottom row shows the result for the next-to-leading jet. The left row and right row correspond to different graviton mass respectively.}
\end{figure}
In Fig.~\ref{fig:dispe}, we display differential cross sections for the transverse momentum $p_T$ of the leading jet and the next-to-leading jet for different center of mass energies and KK graviton masses. We find that the NLO QCD corrections enhance the LO results at both low $p_T$ and high $p_T$. There is a sharply falling in $p_T$ distribution at about half the KK graviton mass, which is called \emph{Jacobian Edge}~\cite{Gordon:1998bn}.  The edge is broadened by the KK graviton width and real corrections at NLO.

\begin{figure}[ht!]
\begin{centering}
  \includegraphics[width=1.0\textwidth]{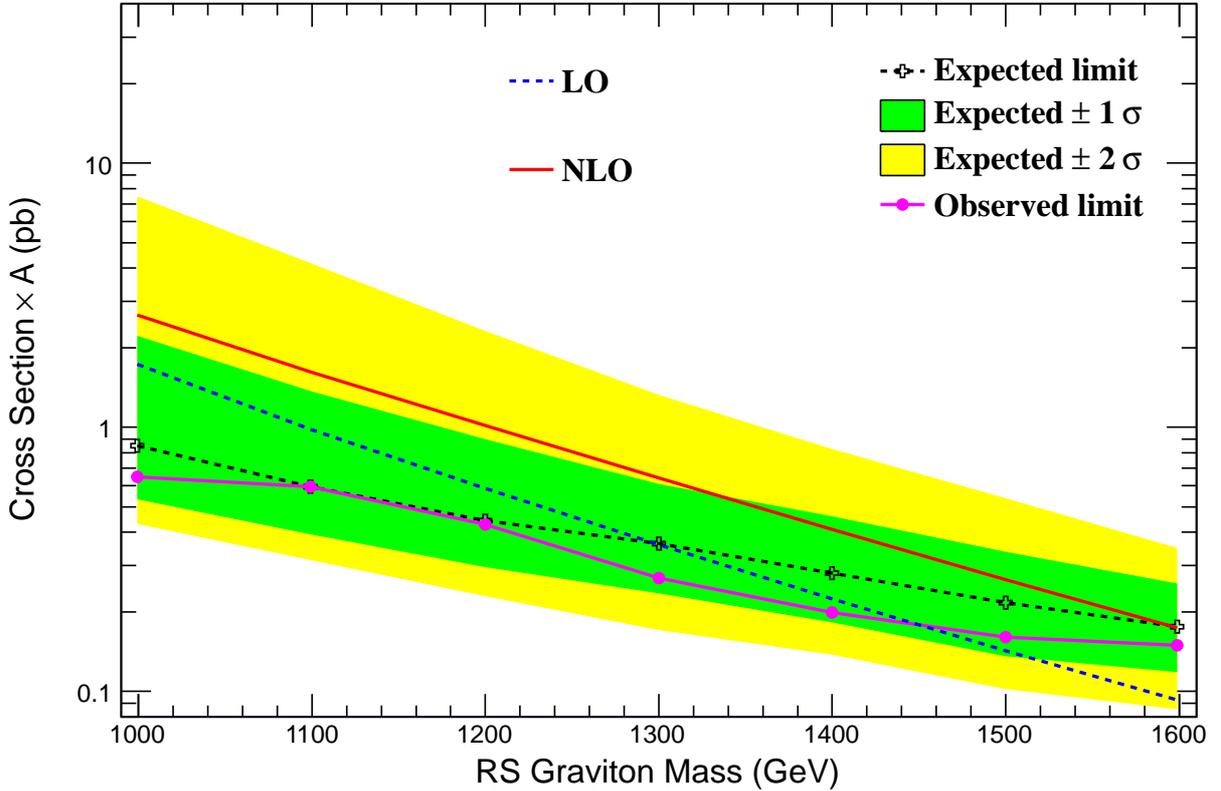}\\
\end{centering}
  \caption{\label{fig:exp}Observed upper limits at $95\%$ CL on $\sigma\times A$ for resonances decaying to dijet final state compared with the expected limits and their variation at the $1\sigma$ and $2\sigma$ levels. }
\end{figure}
\subsection{Signal analysis}\label{ss1}

\begin{figure}[ht!]
\begin{centering}
  \includegraphics[width=0.7\textwidth]{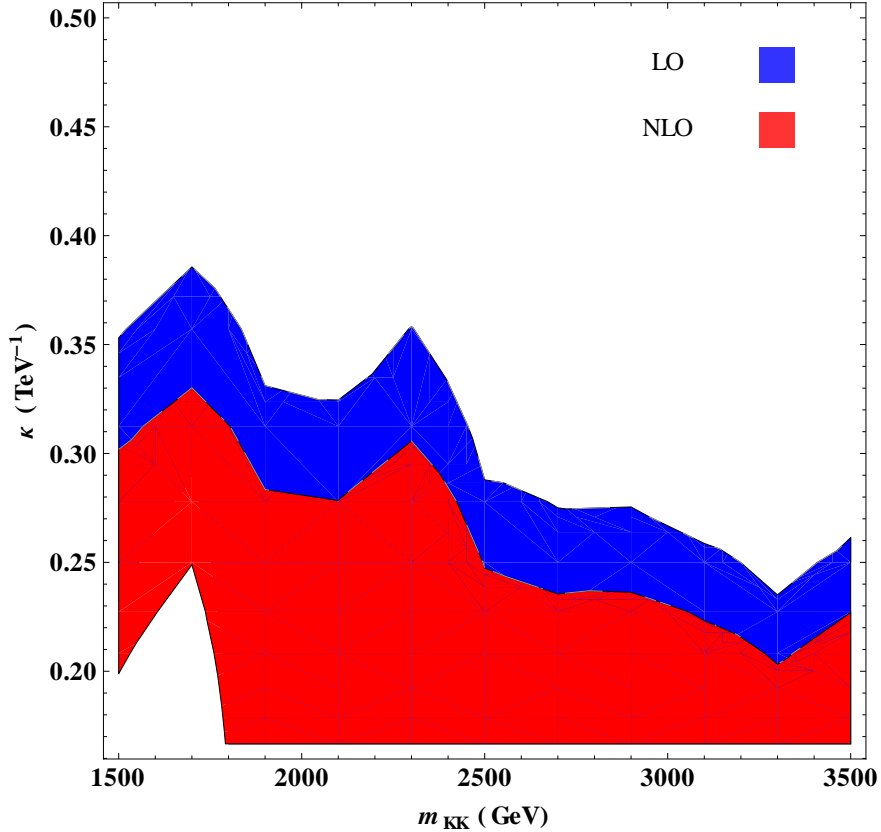}\\
\end{centering}
\caption{\label{fig:para}Allowed parameter space (upper side, 95\% c.l.) for dijet production through KK graviton at the LO and NLO.}
\end{figure}

A search for the KK-graviton has been performed in the dijet mass spectrum by CMS~\cite{Chatrchyan:2013qha}, based on the LO theoretical prediction. Following Ref.~\cite{Chatrchyan:2013qha}, in Fig.~\ref{fig:exp} we present the generic upper limits at the 95\% confidence level for the cross section $\sigma\times A$, where A represents efficiency due to the kinematic requirement of $|\Delta \eta_{jj}|<1.3$ and $m_{jj}>890{\rm~GeV}$. In this subsection, we use CTEQ6M PDF for NLO calculation and CTEQ6L1 PDF for LO calculation. We also assume $k/{\bar{M}_{pl}}=0.1$ for the RS model. Due to the large QCD NLO corrections, the upper limit of excluded mass range of the graviton is promoted from $1.45{\rm~TeV}$ to more than $1.6{\rm~TeV}$.

Fig.~\ref{fig:para} show the allowed parameter space for KK graviton mass and its coupling to SM particles, based on the upper limit for the total cross section in~\cite{Chatrchyan:2013qha}. In our calculation, we consider the coupling region $0.15{\rm~TeV^{-1}}\leq{\kappa}\leq 0.50{\rm~TeV^{-1}}$, and the mass region $1.5{\rm~TeV}\leq m_{KK}\leq 3.5{\rm~TeV}$ same as in the experiment analysis. In Fig.~\ref{fig:para}, the red and blue region corresponds to the 95\% c.l. exclusions at the LO and NLO, respectively. It can be seen from Fig.~\ref{fig:para} that the NLO corrections significantly tighten the allowed parameter space.

\section{CONCLUSION}\label{s6}
In conclusion, we have investigated  dijet production in RS model at the LHC, including QCD NLO corrections to the production and decay of KK graviton. Our results show that the QCD NLO corrections increase the total cross sections by more than $80\%$ and reduce the scale uncertainties. Furthermore, we also explore the distributions for final state dijet invariant mass, jet transverse momentum with QCD NLO accuracy.  Finally, we discuss the constraints on the KK graviton mass and the allowed parameter space of graviton mass and its coupling, based on dijet measurement at the LHC. We find that the upper limit of the KK graviton excluded mass range is promoted from $1.45{\rm~TeV}$ to more than $1.6{\rm~TeV}$ based on our NLO calculations. The allowed parameter space is tightened as well.

\begin{acknowledgments}
We would like to thank Ze Long Liu, Ding Yu Shao and Yan Wang for helpful discussions. This work was supported in part by the National Natural Science Foundation of China under Grants No. 11375013 and No. 11135003, and by the U.S. DOE Early Career Research Award DE-SC0003870 and by Lightner-Sams Foundation.
\end{acknowledgments}

\appendix

\section{Hard non-collinear partonic cross section}\label{s8}
In this appendix we collect the hard non-collinear amplitude square. We use Breit-Wigner approximation and ignore the interference between initial and final state radiation. For simplicity, we define the following invariant variables:
\be
s_{ij}=(p_i+p_j)^2\,.
\ee
For radiations from incoming partons, we have
\beqn
&&\overline{|\mathcal{M}^{real}_{q \bar{q}(g)\rightarrow q \bar{q}}|}^2=-\frac{n_f\pi\alpha_s \kappa^4 s_{34}R(s)}{48(s_{13}+s_{14}-s_{34})(s_{23}+s_{24}-s_{34})(s_{13}+s_{14}+s_{23}+s_{24}-s_{34})}\\\nn
&&\times[-3s_{34}(s_{13}^2+s_{14}^2+s_{23}^2+s_{24}^2)+3s_{34}(s_{13}+s_{14}+s_{23}+s_{s24})(s_{13}^2+s_{14}^2+s_{23}^2+s_{24}^2)\\\nn
&&-4(s_{13}^3 s_{24}+3s_{13}^2 s_{14} s_{23}+3s_{13}s_{24}(s_{14}^2+s_{23}^2)+s_{14}s_{23}(s_{14}^2+s_{23}^2+3s_{24}^2))]\,,
\eeqn
and
\beqn
&&\overline{|\mathcal{M}^{real}_{q \bar{q}(g)\rightarrow g g}|}^2=\frac{4\pi\alpha_s \kappa^4 s_{34}^2[s_{13}^3 s_{23}+s_{13}s_{23}^3+s_{14}s_{24}(s_{14}^2+s_{24}^2)]R(s)}
{9(s_{13}+s_{14}-s_{34})(s_{23}+s_{24}-s_{34})(s_{13}+s_{14}+s_{23}+s_{24}-s_{34})}\,,
\eeqn
and
\beqn
&&\overline{|\mathcal{M}^{real}_{g g(g)\rightarrow q \bar{q}}|}^2=\frac{9n_f\alpha_s\kappa^4 \pi s_{34}^2R(s)}{32(s_{13}+s_{14}-s_{34})(s_{23}+s_{24}-s_{34})(s_{13}+s_{14}+s_{23}+s_{24}-s_{34})}\\\nn
&&\times[2s_{13}^3 s_{14}+2s_{13}s_{14}^3+3s_{13}^2s_{14}s_{23}+s_{14}^3s_{23}+3s_{13}s_{14}s_{23}^2+s_{14}s_{23}^3+s{13}^3s_{24}\\\nn
&&+3s_{13}s_{14}^2s_{24}+3s_{13}^2s_{23}s_{24}+3s_{14}^2s_{23}s_{24}+3s_{13}s_{23}^2s_{24}+2s_{23}^3s_{24}+3s_{13}s_{14}s_{24}^2\\\nn
&&+3s_{14}s_{23}s_{24}^2+s_{13}s_{24}^3+2s_{23}s_{24}^3-(s_{13}+s_{14}+s_{23}+s_{24})^3s_{34}+3(s_{13}+s_{14}+s_{23}+s_{24})^2s_{34}^2\\\nn
&&-4(s_{13}+s_{14}+s_{23}+s_{24})s_{34}^3+2s_{34}^4]\,,
\eeqn
and
\beqn
&&\overline{|\mathcal{M}^{real}_{g g (g)\rightarrow g g}|}^2=\frac{3\pi\alpha_s \kappa^4 s_{34}^2R(s)}{4(s_{13}+s_{14}-s_{34})(s_{23}+s_{24}-s_{34})(s_{13}+s_{14}+s_{23}+s_{24}-s_{34})}\\\nn
&&\times\{(s_{13}^2+s_{13}s_{23}+s_{23}^2)^2-2s_{34}^3(s_{13}+s_{14}+s_{23}+s_{24})+3s_{34}^2[(s_{13}+s_{23})^2+(s_{14}+s_{24})^2]\\\nn
&&-2s_{34}[(s_{13}+s_{23})^3+(s_{14}+s_{24})^3]+(s_{14}^2+s_{14}s_{24}+s_{24}^2)^2+s_{34}^4\}\,.
\eeqn
The other results can be obtained by crossing symmetry.

\bibliography{tmp}

\begin{thebibliography}{12}
\expandafter\ifx\csname bibnamefont\endcsname\relax
  \def\bibnamefont#1{#1}\fi
\expandafter\ifx\csname bibfnamefont\endcsname\relax
  \def\bibfnamefont#1{#1}\fi

\bibitem{CMS:2012nba}
  CMS Collaboration [CMS Collaboration],
  CMS-PAS-EXO-12-016.


\bibitem{CMS:2012eba}
  CMS Collaboration [CMS Collaboration],
  CMS-PAS-EXO-11-094.


\bibitem{Chatrchyan:2013qha}
  S.~Chatrchyan {\it et al.}  [CMS Collaboration],
  Phys.\ Rev.\ D {\bf 87}, no. 11, 114015 (2013)
  [arXiv:1302.4794 [hep-ex]].




  \bibitem{Randall:1999ee}
  L.~Randall and R.~Sundrum,
  Phys.\ Rev.\ Lett.\  {\bf 83}, 3370 (1999)
  [hep-ph/9905221].

\bibitem{Randall:1999vf}
  L.~Randall and R.~Sundrum,
  Phys.\ Rev.\ Lett.\  {\bf 83}, 4690 (1999)
  [hep-th/9906064].




\bibitem{Hewett:1998sn}
\bibinfo{author}{\bibfnamefont{J.~L.} \bibnamefont{Hewett}},
  \bibinfo{journal}{Phys.\ Rev.\ Lett.} \textbf{\bibinfo{volume}{82}},
  \bibinfo{pages}{4765} (\bibinfo{year}{1999}), \eprint{hep-ph/9811356}.

\bibitem{Davoudiasl:1999jd}
  H.~Davoudiasl, J.~L.~Hewett and T.~G.~Rizzo,
  Phys.\ Rev.\ Lett.\  {\bf 84}, 2080 (2000)
  [hep-ph/9909255].

\bibitem{Han:1998sg}
  T.~Han, J.~D.~Lykken and R.~-J.~Zhang,
  Phys.\ Rev.\ D {\bf 59}, 105006 (1999)
  [hep-ph/9811350].

\bibitem{Mathews:2005bw}
  P.~Mathews, V.~Ravindran and K.~Sridhar,
  JHEP {\bf 0510}, 031 (2005)
  [hep-ph/0506158].


\bibitem{Atwood:1999qd}
  D.~Atwood, S.~Bar-Shalom and A.~Soni,
  Phys.\ Rev.\ D {\bf 62}, 056008 (2000)
  [hep-ph/9911231].



\bibitem{Allanach:2002gn}
  B.~C.~Allanach, K.~Odagiri, M.~J.~Palmer, M.~A.~Parker, A.~Sabetfakhri and B.~R.~Webber,
  JHEP {\bf 0212}, 039 (2002)
  [hep-ph/0211205].




\bibitem{Li:2006yv}
  Q.~Li, C.~S.~Li and L.~L.~Yang,
  Phys.\ Rev.\ D {\bf 74}, 056002 (2006)
  [hep-ph/0606045].

\bibitem{Mathews:2004xp}
  P.~Mathews, V.~Ravindran, K.~Sridhar and W.~L.~van Neerven,
  Nucl.\ Phys.\ B {\bf 713}, 333 (2005)
  [hep-ph/0411018].

\bibitem{Kumar:2006id}
  M.~C.~Kumar, P.~Mathews and V.~Ravindran,
  Eur.\ Phys.\ J.\ C {\bf 49}, 599 (2007)
  [hep-ph/0604135].

\bibitem{Kumar:2009nn}
  M.~C.~Kumar, P.~Mathews, V.~Ravindran and A.~Tripathi,
  Nucl.\ Phys.\ B {\bf 818}, 28 (2009)
  [arXiv:0902.4894 [hep-ph]].



\bibitem{Kumar:2008pk}
  M.~C.~Kumar, P.~Mathews, V.~Ravindran and A.~Tripathi,
  Phys.\ Lett.\ B {\bf 672}, 45 (2009)
  [arXiv:0811.1670 [hep-ph]].

\bibitem{Agarwal:2009xr}
  N.~Agarwal, V.~Ravindran, V.~K.~Tiwari and A.~Tripathi,
  Nucl.\ Phys.\ B {\bf 830}, 248 (2010)
  [arXiv:0909.2651 [hep-ph]].

\bibitem{Agarwal:2009zg}
  N.~Agarwal, V.~Ravindran, V.~K.~Tiwari and A.~Tripathi,
  Phys.\ Lett.\ B {\bf 686}, 244 (2010)
  [arXiv:0910.1551 [hep-ph]].

\bibitem{Agarwal:2010sp}
  N.~Agarwal, V.~Ravindran, V.~K.~Tiwari and A.~Tripathi,
  Phys.\ Rev.\ D {\bf 82}, 036001 (2010)
  [arXiv:1003.5450 [hep-ph]].

\bibitem{Agarwal:2010sn}
  N.~Agarwal, V.~Ravindran, V.~K.~Tiwari and A.~Tripathi,
  Phys.\ Lett.\ B {\bf 690}, 390 (2010)
  [arXiv:1003.5445 [hep-ph]].

\bibitem{Chen:2014oha}
   C.~Y.~Chen, H.~Davoudiasl and D.~Kim,
   Phys.\ Rev.\ D {\bf 89}, 096007 (2014)
   [arXiv:1403.3399 [hep-ph]].

\bibitem{Gao:2010bb}
  J.~Gao, C.~S.~Li, B.~H.~Li, H.~X.~Zhu and  C.~-P.~Yuan,
  Phys.\ Rev.\ D {\bf 82}, 014020 (2010)
  [arXiv:1004.0876 [hep-ph]].


\bibitem{Cao:2004ap}
  Q.~H.~Cao, R.~Schwienhorst and C.-P.~Yuan,
  Phys.\ Rev.\ D {\bf 71}, 054023 (2005)
  [hep-ph/0409040].



\bibitem{Fadin:1993kt}
  V.~S.~Fadin, V.~A.~Khoze and A.~D.~Martin,
  Phys.\ Lett.\ B {\bf 320}, 141 (1994)
  [hep-ph/9309234].

\bibitem{Fadin:1993dz}
  V.~S.~Fadin, V.~A.~Khoze and A.~D.~Martin,
  Phys.\ Rev.\ D {\bf 49}, 2247 (1994).

\bibitem{Melnikov:1993np}
  K.~Melnikov and O.~I.~Yakovlev,
  Phys.\ Lett.\ B {\bf 324}, 217 (1994)
  [hep-ph/9302311].


\bibitem{Beenakker:1988jr}
  W.~Beenakker and A.~Denner,
  Nucl.\ Phys.\ B {\bf 338}, 349 (1990).

\bibitem{Harris:2001sx}
  B.~W.~Harris and J.~F.~Owens,
  Phys.\ Rev.\ D {\bf 65}, 094032 (2002)
  [hep-ph/0102128].

\bibitem{Altarelli:1977zs}
  G.~Altarelli and G.~Parisi,
  Nucl.\ Phys.\ B {\bf 126}, 298 (1977).

\bibitem{altarelli}
   G.~Altarelli, R.~K.~Ellis, G.~Martinelli,
   Nucl. Phys. B 157 (1979) 461; J.~C.~Collins, D.~E.~Soper and
   G.~Sterman, in: {\it Perturbative Quantum Chromodynamics},
   ed. A.H. Mueller (World Scientific, 1989).

\bibitem{Gorishny:1990zu}
  S.~G.~Gorishny, A.~L.~Kataev, S.~A.~Larin and L.~R.~Surguladze,
  Mod.\ Phys.\ Lett.\ A {\bf 5}, 2703 (1990).
  S.~G.~Gorishny, A.~L.~Kataev, S.~A.~Larin and L.~R.~Surguladze,
  Phys.\ Rev.\ D {\bf 43}, 1633 (1991).
  A.~Djouadi, M.~Spira and P.~M.~Zerwas,
  Z.\ Phys.\ C {\bf 70}, 427 (1996)
  [hep-ph/9511344].
  A.~Djouadi, J.~Kalinowski and M.~Spira,
  Comput.\ Phys.\ Commun.\  {\bf 108}, 56 (1998)
  [hep-ph/9704448].
  M.~Spira,
  Fortsch.\ Phys.\  {\bf 46}, 203 (1998)
  [hep-ph/9705337].

\bibitem{Pumplin:2002vw}
  J.~Pumplin, D.~R.~Stump, J.~Huston, H.~L.~Lai, P.~M.~Nadolsky and W.~K.~Tung,
  JHEP {\bf 0207}, 012 (2002)
  [hep-ph/0201195].

\bibitem{Cacciari:2008gp}
  M.~Cacciari, G.~P.~Salam and G.~Soyez,
  JHEP {\bf 0804}, 063 (2008)
  [arXiv:0802.1189 [hep-ph]].



\bibitem{Cacciari:2011ma}
  M.~Cacciari, G.~P.~Salam and G.~Soyez,
  Eur.\ Phys.\ J.\ C {\bf 72}, 1896 (2012)
  [arXiv:1111.6097 [hep-ph]].

\bibitem{Aad:2009wy}
  G.~Aad {\it et al.}  [ATLAS Collaboration],
  arXiv:0901.0512 [hep-ex].

\bibitem{Gordon:1998bn}
  A.~S.~Gordon,
  FERMILAB-THESIS-1998-10.



\end{thebibliography}

\end{document}